\documentclass[12pt]{article}
\usepackage{chet}
\usepackage{booktabs}
\usepackage{amsmath,amssymb}
\usepackage{graphicx,url}
\usepackage{color}
\usepackage{setspace}
 
\usepackage{epsfig}

\def\Or[#1]{{\text{O}}\left({#1}\right)}
\def\dotl[#1,#2]{\left\langle #1, #2 \right\rangle}
\def\dotlb[#1,#2]{[ #1, #2 ]}
\def\dotp[#1,#2]{(#1) \cdot (#2)}
\def\aff[#1,#2]{\hat{#1}(#2)}
\def\n4sym{{\cal N}=4 SYM}
\def\>{\rangle}
\def\<{\langle}
\def\weight[#1,#2,#3]{\{(#1),#2,#3\}}
\def\ads[#1]{$\text{AdS}_{#1}$}

\newcommand{\ba}{\begin{eqnarray}}
\newcommand{\ea}{\end{eqnarray}}
\newcommand{\be}{\begin{eqnarray}}
\newcommand{\ee}{\end{eqnarray}}
\newcommand{\bq}{\begin{equation}}
\newcommand{\eq}{\end{equation}}
\newcommand{\benn}{\begin{equation*}}
\newcommand{\eenn}{\end{equation*}}
\newcommand{\bi}{\begin{itemize}}  
\newcommand{\ei}{\end{itemize}}

\newcommand{\CF}{{\cal F}}

\newcommand{\CO}{{\cal O}}

\newcommand{\nn}{\nonumber}
\renewcommand{\Im}{\text{Im}}

\newcommand\oo\infty
\newcommand\s\sigma
\newcommand\de\delta
\newcommand\De\Delta

\newcommand\f\phi
\newcommand\g\gamma
\newcommand\x\times

\newcommand\R{\mathbb{R}}

\usepackage{tikz}

\begin{document}
\usetikzlibrary{calc,decorations.markings,decorations.pathmorphing}

\title{Closure of the Operator Product Expansion in the Non-Unitary Bootstrap}
\author{Ilya Esterlis$^1$, A. Liam Fitzpatrick$^2$, David M.~Ramirez$^1$}
\affiliation{
{\it $^1$ Stanford Institute for Theoretical Physics, Stanford University, \\
Via Pueblo, Stanford, CA 94305, U.S.A.} \\
{\it $^2$  Dept.\ of Physics, Boston University, Boston, MA 02215} \\

}
\abstract{ 
We use the numerical conformal bootstrap in two dimensions to search for finite, closed sub-algebras of the operator product expansion (OPE), without assuming unitarity.  We find the minimal models as special cases, as well as additional lines of solutions that can be understood in the Coulomb gas formalism. All the solutions we find that contain the vacuum in the operator algebra are cases where the external operators of the bootstrap equation are degenerate operators, and we argue that this follows analytically from the expressions in arXiv:1202.4698 for the crossing matrices of Virasoro conformal blocks.  Our numerical analysis is a special case of the ``Gliozzi'' bootstrap method, and provides a simpler setting in which to study technical challenges with the method.  

 In the supplementary material, we provide a Mathematica notebook that automates the calculation of the crossing matrices and OPE coefficients for degenerate operators using the formulae of Dotsenko and Fateev.
}

\maketitle

\tableofcontents

\flushbottom

\section{Introduction and Summary}

Through the conformal bootstrap, it is possible at least in principle to ask precisely what is the full space of Conformal Field Theories (CFTs).  Answering this question in full generality is beyond the ability of currently available techniques, but for special classes of CFTs it sometimes does become a tractable problem.  The infinite conformal symmetry of two-dimensional CFTs makes them a natural place to start.  In two-dimensions, a sufficient set of consistency conditions that a CFT must satisfy is modular invariance of zero- and one-point functions on the torus together with crossing symmetry of four-point functions on the sphere \cite{Moore:1988uz}.\footnote{Through the introduction of ``twist operators,'' the first two of these three cases can be formulated as special cases of the third \cite{Dijkgraaf:1989hb,Lunin:2000yv}.}  However, these constraints still involve an infinite set of data and thus searching for all solutions to the constraints is intractable.  One strategy is to look for theories where only a finite amount of such `data' is non-trivial, in which case finding solutions can become tractable.  This strategy famously leads to the `minimal models' \cite{Belavin:1984vu}, theories with only a finite number of (Virasoro) primary operators.  Using modular invariance, such theories have been completely classified.  While extremely fruitful, there is still clearly qualitative behavior allowed in general CFTs that does not arise in minimal models (in particular, behavior associated with large central charge gravity duals).

A natural generalization of this strategy is not to use the full set of constraints of the theory, but instead to see what can be obtained from just the constraint of crossing symmetry of a small number of four-point functions.  Demanding that {\it these} depend on only a finite set of data is apparently a much weaker condition than demanding it of the full theory.  That is, one can hope that there exist CFTs that have an infinite number of primary operators, but which have a finite sub-algebra of operators that closes under the OPE. The most drastic such constraint would be to demand closure with just a single scalar operator $\phi$ (in addition to the identity operator $1$):
\be
[\phi] \times [\phi] =[1],
\label{eq:ansatz0}
\ee
where $[\phi]$ denotes the entire Verma module associated to the primary operator $\phi$.\footnote{ Of course, the vacuum $[\phi]=[1]$ is an example of such an operator algebra, but one that would be considered trivial.}  Generalizing only slightly, without introducing any additional operators, we can relax the above constraint to allow $[\phi]$ to appear in its own OPE:
\be
[\phi] \times [\phi] = [1] + [\phi].
\label{eq:ansatz1}
\ee
Searching for such operator algebras is a simple problem in the conformal bootstrap, and can be solved numerically.  In particular, it is simple enough that it does not require any assumption about unitarity.  

Perhaps surprisingly, we find that in all solutions to this equation, the conformal weight of $\phi$ and the central charge $c$ of the Virasoro algebra are those of one or another minimal model.  That is, all the solutions we find numerically by imposing crossing symmetry are covered by the 
minimal model formulae for the central charge $c$ and weights $h_\phi = h_{r,s}$, 
	\begin{align}
	c&=1-\frac{6(p-p')^2}{pp'} , 	\qquad h_{r,s} =\frac{(pr-p's)^2-(p-p')^2}{4pp'}, 
\end{align}
with
\be
\left[\phi\right] \times \left[\phi\right] = \left[1\right] &:& (r,s) = (1,1) \textrm{ or } (1, p-1), \nn\\
\left[\phi\right] \times \left[\phi\right] = \left[1\right] + \left[ \phi \right]&:& (r,s) = (1,2), \qquad \qquad p'=2,  p=5, \nn\\
&& (r,s) = (1,p-2), \qquad p'=2,3,4, p=5, \nn\\
&& (r,s) = (2, p-1), \qquad p'=5, p>5 ,
\label{eq:opetrunc01}
\ee
up to dualities $(r,s) \cong (p'-r, p-s)$, 
and we have taken $p>p'$ without loss of generality.\footnote{In appendix \ref{app:minmodtrunc}, we review the truncation of the OPE algebra for these minimal model operators. } 
Thus, this ``weaker'' condition is in fact enough to essentially imply the much stronger conditions mentioned above.

To explore more widely, we also consider the case where an extra operator, not necessarily the same as $\phi$, may appear in the OPE. That is, we demand that the four-point function $\< \phi(z_1) \phi(z_2) \phi(z_3) \phi(z_4)\>$ obeys crossing with only the following OPE content:
\be
[\phi] \times [\phi] = [1]+ [\epsilon].
\label{eq:ansatz2}
\ee
A solution to this equation is not necessarily a full-fledged solution to closure of the Operator Product Expansion, because the operator product of $\phi$ with $\epsilon$ (or $\epsilon$ with $\epsilon$) may produce yet additional operators.  Nevertheless, it can be solved, and we find two classes of solutions that are not technically minimal models. The first class of solutions is just the one described in \cite{Liendo:2012hy}, where $\phi$ is a state with null descendants at level 2.  Such solutions are quite similar in spirit to minimal models, but are known to imply an infinite number of operators in the full theory by the constraints of modular invariance.  More generally, they are part of the class of degenerate operators parameterized by 
\be
\left[\phi\right] \times \left[\phi\right] = \left[1\right] + \left[ \epsilon \right]&:& (r,s) = (1,2),(2,1), \qquad \qquad p', p = \textrm{any} \nn\\
&& \textrm{ or } (1, p-2), (2,p-1), \qquad  p',p \in \mathbb{Z},
\label{eq:opetrunc2}
\ee
again up to dualities and taking $p>p'$.  The $(r,s) = (1,2)$ or $(2,1)$ operators do not require $p',p$ to be coprime integers or even to be well-defined; the point is that for any value of $c$, these operators have null descendants, but are not necessarily part of a unitary, rational CFT.

The second class is more unusual.  In this class, the vacuum block actually decouples, and more precisely one obtains the OPE
\be
[\phi] \times [\phi] = [\epsilon].
\ee
This OPE is possible when the following relations hold:
\be
h_\epsilon &=& \frac{4}{3} h_\phi, \nn\\
c &=& 32  h_\phi + 1.
\ee
One can think of $\phi$ in this case as a degenerate operator with $r=s=\frac{1}{2}$, and we will see that in the Coulomb gas formalism this choice of $(r,s)$ leads to a particularly simple form for the $\epsilon$ Virasoro conformal block, $ \propto (z(1-z))^{-2h_\phi/3}$.  The decoupling of the vacuum block implies that the state $\phi$ has vanishing norm.  Alternatively, one can consider starting with a crossing-symmetric four-point function of $\phi$ that {\it does} contain the vacuum block, and then adding the $\epsilon$ block with an OPE coefficient $C_{\phi \phi \epsilon}$ in order to generate a continuous line of solutions to the crossing equation at fixed $h_\phi$ and $c=32h_\phi+1$ but arbitrary $C_{\phi \phi \epsilon}$. Such lines are interesting from the point of view that they represent an ambiguity in the solution of the bootstrap equation even after one specifies the spectrum of conformal blocks appearing in the $\< \phi\phi\phi\phi\>$ four-point function.\footnote{This kind of ambiguity was discussed in \cite{Gliozzi:2014jsa} in the context of global conformal invariance and $O(n)$ models.  The difference in our ambiguity is that it allows one to dial the OPE coefficient of a single Virasoro conformal block, without affecting any of the others.}

We perform most of our analysis with the Gliozzi bootstrap method \cite{Gliozzi:2013ysa,Gliozzi:2014jsa}, which looks for points in parameter space at which a certain rectangular matrix has a nontrivial kernel. The condition to have a nontrivial kernel can be phrased in terms of the simultaneous vanishing of sub-determinants of this matrix. An alternative way of stating this is that the matrix must have at least one vanishing singular value. In certain cases, we found the latter statement to be more useful.  There were two primary reasons for this. First, because singular values are nonnegative, looking for vanishing singular values becomes a minimization problem. Such problems are numerically much more robust than root finding. Second, we found the singular value method avoids subtleties associated with the determinant method. We discuss these issues and illustrate the advantage of the singular value approach in more detail at the end of section 3.

 In the final section of the paper, we seek to give at least a partial analytic proof of the numeric results.  To do this, we turn to remarkable results on the ``crossing matrix'' $F_{\alpha_s, \alpha_t} [{ \alpha_2 \ \alpha_3 \atop \alpha_1 \ \alpha_4}]$ for Virasoro conformal blocks \cite{Teschner1,Teschner2,Teschner3}.  This is the matrix that describes the decomposition of (the holomorphic part of) a Virasoro conformal block in one channel in terms of Virasoro conformal blocks in another channel:
 \be
 \label{eq:holomorphic-fusion-matrix}
 {\cal F}(\alpha_s, \alpha_i, c,z) = \int d\alpha_t F_{\alpha_s, \alpha_t} \left[{ \alpha_2 \ \alpha_3 \atop \alpha_1 \ \alpha_4}\right] {\cal F}(\alpha_t, \alpha_i, c,1-z).
 \ee
The crossing matrix is an efficient way to encapsulate the problem of finding correlators that satisfy the bootstrap equation. For instance, consider a four-point function with all external operators equal, $\alpha_i = \alpha$.  Then, if one decomposes such a correlator $G(z)$ in a basis of conformal blocks,\footnote{Here we suppress the antiholomorphic piece for notational simplicity but it will be included in the subsequent analysis.}
\be
G(z) = \int d \alpha_s P_{\alpha_s} {\cal F}(\alpha_s, \alpha, c,z) ,
\ee
then $P_{\alpha_s}$ is simply an eigenvector of $F_{\alpha_s, \alpha_t} \left[{ \alpha \ \alpha \atop \alpha \ \alpha}\right] $ with eigenvalue 1:
\be
\int d\alpha_s P_{\alpha_s}F_{\alpha_s, \alpha_t} \left[{ \alpha \ \alpha \atop \alpha \ \alpha}\right]  = P_{\alpha_t}.
\ee
We are interested in theories with a discrete spectrum, in which case $P_{\alpha_s}$ is a sum over $\delta$ functions as a function of $\alpha_s$, and consequently $F_{\alpha_s, \alpha_t} \left[{ \alpha \ \alpha \atop \alpha \ \alpha}\right]$ must be as well when evaluated on the values of $\alpha_t$ that appear in the decomposition of $G(z)$.  In the case where the dimensions and central charge of the theory take the values of minimal model theories, this can be seen explicitly, and in fact only a sum over a finite number of $\delta$ functions appears.  This provides an efficient way of obtaining correlators in minimal models, since the problem is reduced to finding the eigenvalues of a finite-dimensional matrix.   More generally, the constraints that the OPE satisfy (\ref{eq:ansatz1}) or (\ref{eq:ansatz2}) imply that $F_{\alpha_s, \alpha_t} \left[{ \alpha \ \alpha \atop \alpha \ \alpha}\right]$ reduce to a sum over a finite number of $\delta$ functions, and this combined with the formulae for $F$  give strong constraints on the spectrum of operators.  We will see in section \ref{sec:tesch} that these constraints make it extremely hard, if not impossible, to satisfy (\ref{eq:ansatz1}) for any values of operator dimension other than those in minimal models.

In the case of minimal models, of course, the crossing matrices $F$ do become finite-dimensional matrices.  Explicit formulae for them are known from the work of \cite{Dotsenko:1984nm,Dotsenko:1984ad}, and these have been useful to us both for providing consistency checks, as well as for providing an efficient method for exploring minimal models in the context of crossing symmetry.  In the supplementary material, we provide a brief Mathematica notebook that evaluates the formulae from \cite{Dotsenko:1984nm,Dotsenko:1984ad} for the crossing matrices in minimal models, as well as for their OPE coefficients.

\section{Bootstrap Review}

The  method we use will be analogous to that proposed in \cite{Gliozzi:2013ysa}. There, the authors worked with the global conformal algebra. Here we first utilize the full Virasoro algebra and later restrict to the global algebra. The method is especially well suited to address our question as it does not require unitarity as an input (in practice this means one does not demand positivity of squared OPE coefficients), and so can be expected to apply for non-unitary as  well as unitary theories. 

Consider the four point function of identical, primary scalar operators, $\langle\phi(x_1)\phi(x_2)\phi(x_3)\phi(x_4)\rangle$, with conformal weights  $h = \bar{h}$. Here $x_i$ denotes the pair $(z_i,\bar z_i)$. Global conformal symmetry constrains the four point function to have the form
	\begin{equation}
	\langle \phi(x_1)\ldots \phi(x_4)\rangle=\frac{1}{|z_{12}|^{4h} |z_{34}|^{4h}} f(\eta,\bar \eta),
	\end{equation}
where
	\begin{equation}
	z_{ij}=z_i-z_j, \quad \eta=\frac{z_{12}z_{34}}{z_{13}z_{24}}.
	\end{equation}
Global conformal symmetry further allows us to put $z_1=\infty$, $z_2=1$, $z_4=0$, and $z_3=z$. The conformally-invariant cross ratio becomes $\eta=z$ and we have
	\begin{equation}
	\langle \phi(\infty)\phi(1)\phi(z,\bar z) \phi(0)\rangle\equiv \lim_{z_1,\bar z_1\rightarrow\infty}z_1^{2h}\bar z_1^{2h}\langle \phi(z_1,\bar z_1)\phi(1)\phi(z,\bar z) \phi(0)\rangle=G(z,\bar z).
	\end{equation}
The function $G(z,\bar z )$ has the conformal block decomposition
	\begin{equation}
 	G(z,\bar z)=\sum_p a_p\mathcal{F}(c,h_p,h,z)\bar{\mathcal{F}}(c,\bar h_p,h,\bar z),
  	\label{eq:vir+expand}
	\end{equation}
where the sum on $p$ is a sum over Virasoro primaries,  $a_p$ are the squared OPE coefficients, and the functions $\mathcal F$ are Virasoro conformal blocks.\footnote{In practice we will use Zamolodchikov's recursion relation \cite{ZamolodchikovRecursion,Zamolodchikovq} for the blocks, using a modification of the Mathematica code provided in \cite{HartmanLargeC}.}
The above equation is an expansion in the $s$-channel, $z\rightarrow 0$. Expanding instead in the $t$-channel, $z\rightarrow 1$, and demanding equality to the $s$-channel expression gives the crossing condition $G(z,\bar z)=G(1-z,1-\bar z)$. Using the expansion \eqref{eq:vir+expand}, we write this as a sum rule:
	\begin{equation}
	\sum_p a_p[\mathcal{F}(c,h_p,h,z)\bar{\mathcal{F}}(c,\bar h_p,h,\bar z)-\mathcal{F}(c,h_p,h,1-z)\bar{\mathcal{F}}(c,\bar h_p,h,1-\bar z)]=0.
	\label{eq:homog+eqn}
	\end{equation}
Expanding this about the point $z=\bar z=1/2$ gives an infinite set of homogeneous equations
	\begin{equation}
	\sum_p a_p g^{(m,n)}_{h,\bar h}= 0,
	 \label{eq:vir+gli+eqn}
	\end{equation}
 where
	\begin{equation}
	g^{(m,n)}_{h,\bar h}=\left.\partial_z^m\partial_{\bar z}^n\left[\mathcal{F}(c,h_p,h,z)\bar{\mathcal{F}}(c,\bar h_p,h,\bar z)-\mathcal{F}(c,h_p,h,1-z)\bar{\mathcal{F}}(c,\bar h_p,h,1-\bar z)\right]\right|_{z=\bar z=1/2},
	\label{eq:blockderiv}
	\end{equation}
and  without loss of generality we can restrict to $ m> n \ge 0$  with  $m+n$ odd. We find it more robust to work with derivatives of the blocks directly, eq.~(\ref{eq:blockderiv}), than derivatives of them normalized by the vacuum block, which are often used.\footnote{ One reason is that near the minimal models, individual blocks' contribution $\CF(z)-\CF(1-z)$ divided by the vacuum block's contribution often becomes a constant, and therefore all of its derivatives to vanish.  The reason for this is fairly easy to understand in terms of the crossing matrices $F=F_{\alpha_s, \alpha_t}\left[ \begin{array}{cc} \alpha_1 & \alpha_2 \\ \alpha_3 & \alpha_4 \end{array} \right]$ that we discuss in more detail in section \ref{sec:tesch}. The point is that when one of the external operators is a degenerate operator, the crossing matrix is finite-dimensional and squares to 1, $F^2= \mathbf{1}$.  Solutions of the crossing matrix are eigenvectors with eigenvalue 1, and since $F^2=\mathbf{1}$, all of its eigenvalues are either 1 or $-1$.  Therefore, it will generally have not just a  unique solution, but a linear subspace of solutions, namely the space generated by the eigenvalue-1 eigenvectors. In the case of the null vector $(r,s) = (1,3)$, for example, there are only three operators in its OPE, which we can call $[\CO_{1,1}]$ (the vacuum), $[\CO_{1,3}]$, and $[\CO_{1,5}]$, and $F$ has two eigenvalues equal to 1.  As a result, the space of solutions is one-dimensional, and one can without loss of generality set the coefficient of the $\CO_{1,5}$ block to zero and still get a solution to crossing.  But this means that  the $\CO_{1,3}$ contribution to the crossing equation is a multiple of the $\CO_{1,1}$ contribution, and therefore their ratio is constant. }
The approach of \cite{Gliozzi:2013ysa} comes from the observation that, for an OPE including $N$ primaries, \eqref{eq:vir+gli+eqn} will have a nontrivial solution if and only if all the minors of order $N$ of the matrix $g^{(m,n)}_{h,\bar h}$ are nonvanishing. Taking $M\geq N$ derivatives then gives a set of $\kappa=\begin{pmatrix}M \\ N \end{pmatrix}$ equations. 

The OPE \eqref{eq:ansatz1} corresponds to $N = 2$ Virasoro primaries with the central charge $c$ and conformal weight $h$ of operator $\phi$ as the only free parameters. Thus, taking $M > 2$, we obtain an over-constrained system of $\kappa$ equations for $c$ and $h$. Solutions to this system give four point functions consistent with crossing symmetry, containing the single primary operator $\phi$. We stress there are no unitarity constraints imposed on either $c$ or $h$, so this method should find both unitary and non-unitary crossing-symmetric four point functions.

Of course, in principle we  did  not need to restrict to the extremely small sub-algebras we consider here; any finite size would do.  With $N$ operators in the algebra, there are $\CO(N^3)$ free parameters to solve for, so the size of the parameter space becomes much larger and numerically the problem would appear to be much more challenging.   However, one of the main points of \cite{Gliozzi:2013ysa} was that one can formulate the problem in terms of finding the solution to a non-linear function of the operator dimensions, which in this case would be only $\CO(N)$ free parameters.   It seems likely that studying larger finite closed sub-algebras may be an ideal setup to explore in even greater detail how to reduce systematic uncertainties in the methods of \cite{Gliozzi:2013ysa} more generally.\footnote{Some comments along these lines appear in \cite{Gliozzi:2016cmg}. }

\section{Results}

\subsection{$[\phi] \times [\phi] = [1]$}

As a warm-up we consider the OPE \eqref{eq:ansatz0}, in which the operator $\phi$ squares to the identity. In this case \eqref{eq:vir+gli+eqn} simplifies to 
\be
\left.\partial_z^m\partial_{\bar z}^n \mathcal{F}_{\rm vac}(c,h,z)\bar{\mathcal{F}}_{\rm vac}(c,h,\bar z)\right|_{z=\bar z=1/2} = 0,
\ee
with $ m + n$ odd.  Note that this equation can be factored into the form
\be
\partial_z^m \mathcal{F}_{\rm vac}(c,h,z)_{z=1/2} = 0 &\textrm{ or }& \partial_z^n \mathcal{F}_{\rm vac}(c,h,z)_{z=1/2} = 0
\ee
for all $m+n$ odd.   This constraint immediately implies that either all even derivatives vanish or all odd derivatives vanish.  The reason is that if even a single  $\partial_z^m$ derivative with $m$ odd does not vanish, then one can make $m+n$ odd by taking $n$ to be any even number, and so all even derivatives must vanish.  Similarly, if even a single $\partial_z^m$ derivative with $m$ even does not vanish, then all the odd derivatives must vanish.  
In practice, we have found that all solutions to crossing with $c>0$ have vanishing odd derivatives, and all solutions with $c<0$ have vanishing even derivatives, though we do not have a simple explanation for this fact.

\begin{figure}[t!]
\begin{center}
\includegraphics[width=0.45\linewidth]{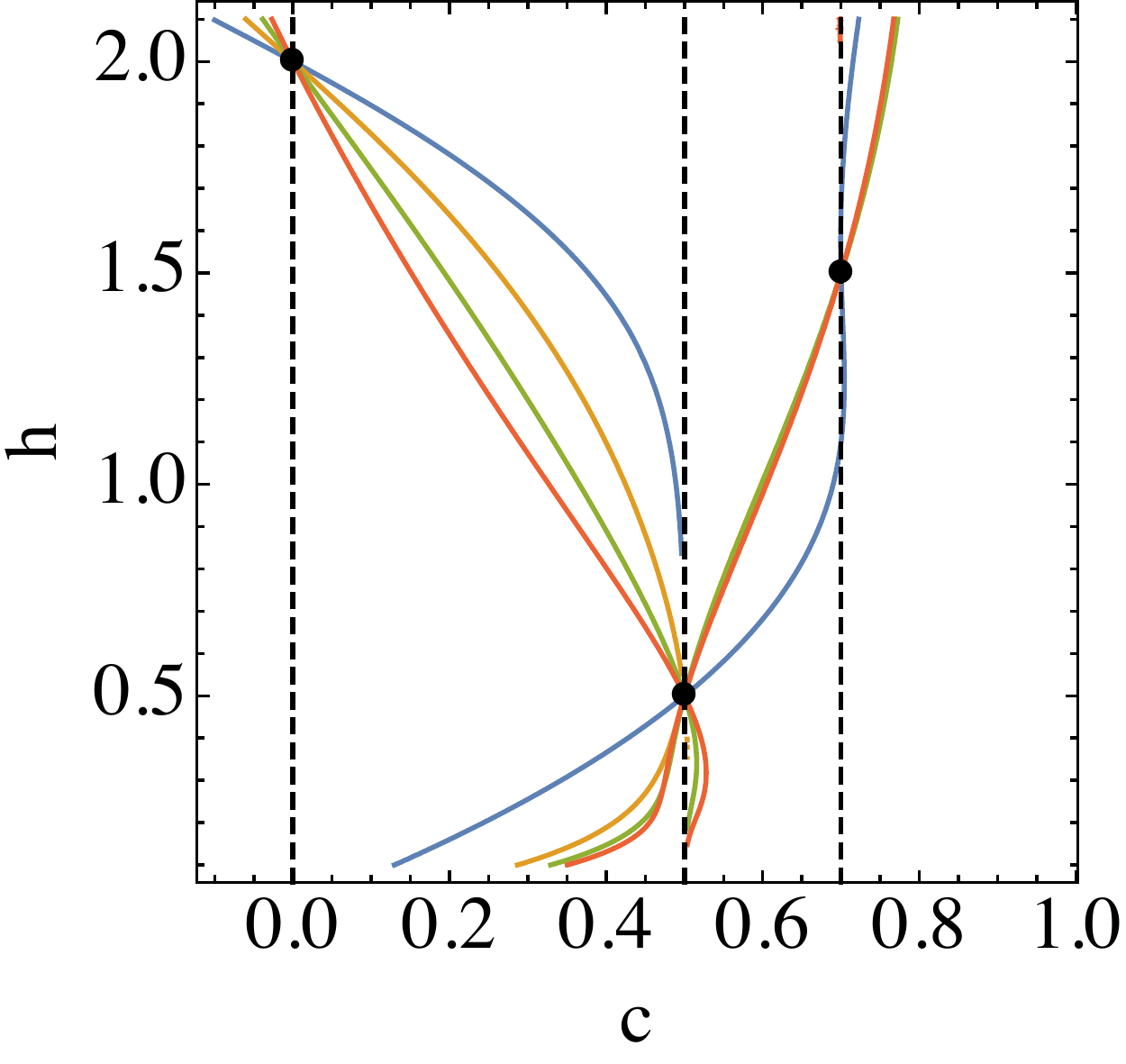}
\end{center}
\caption{ A plot of the zero contours of the first, third, fifth, and seventh derivatives of the vacuum block. Black dots indicate minimal model operators with OPE closure $[\phi]\times[\phi] = [1]$. For reference, the values are $(c,h)=(0,2),(1/2,1/2),(7/10,3/2)$. Vertical lines are $c=0,1/2,7/10$. }
\label{fig:vac+closure0c1}
\end{figure}

\begin{figure}[h]
\begin{center}
\includegraphics[width=0.45\linewidth]{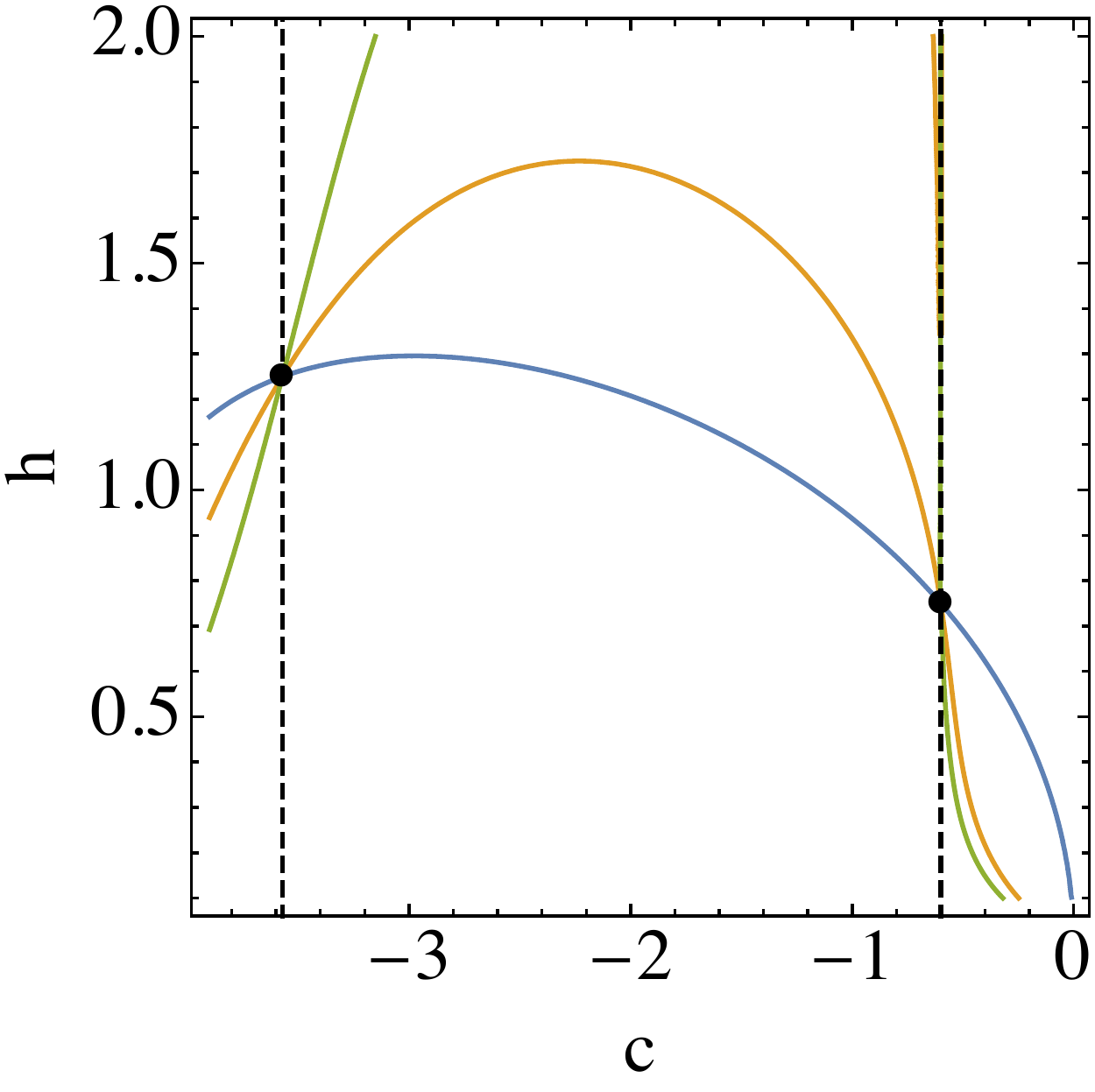}
\end{center}
\caption{ Same as figure \ref{fig:vac+closure0c1}, except in the region $c<0$, and showing the zero contours of the zeroth, second, and fourth derivatives of the vacuum block. 
 Values are $(c,h)=(-3/5,3/4),(-25/7,5/4)$. Vertical lines are $c=-3/5,-27/5$.}
\label{fig:vac+closure-4c0}
\end{figure}

We look for solutions in the region
    \be
    \mathcal R = \{(c,h) : -4 \leq c \leq 1, ~ 0 \leq h \leq 2  \} \label{eq:region}
    \ee
and take $M\leq 7$. Contours of vanishing derivatives, as functions of $c$ and $h$, are shown in figures \ref{fig:vac+closure0c1} and \ref{fig:vac+closure-4c0}. Points where all contours intersect are putative solutions to \eqref{eq:vir+gli+eqn}. In this region we find there are \textit{no} solutions other than the known minimal models. 


\subsection{$[\phi] \times [\phi] = [1]+[\phi]$}

\begin{figure}[ht!]
\begin{center}
\includegraphics[width=0.45\linewidth]{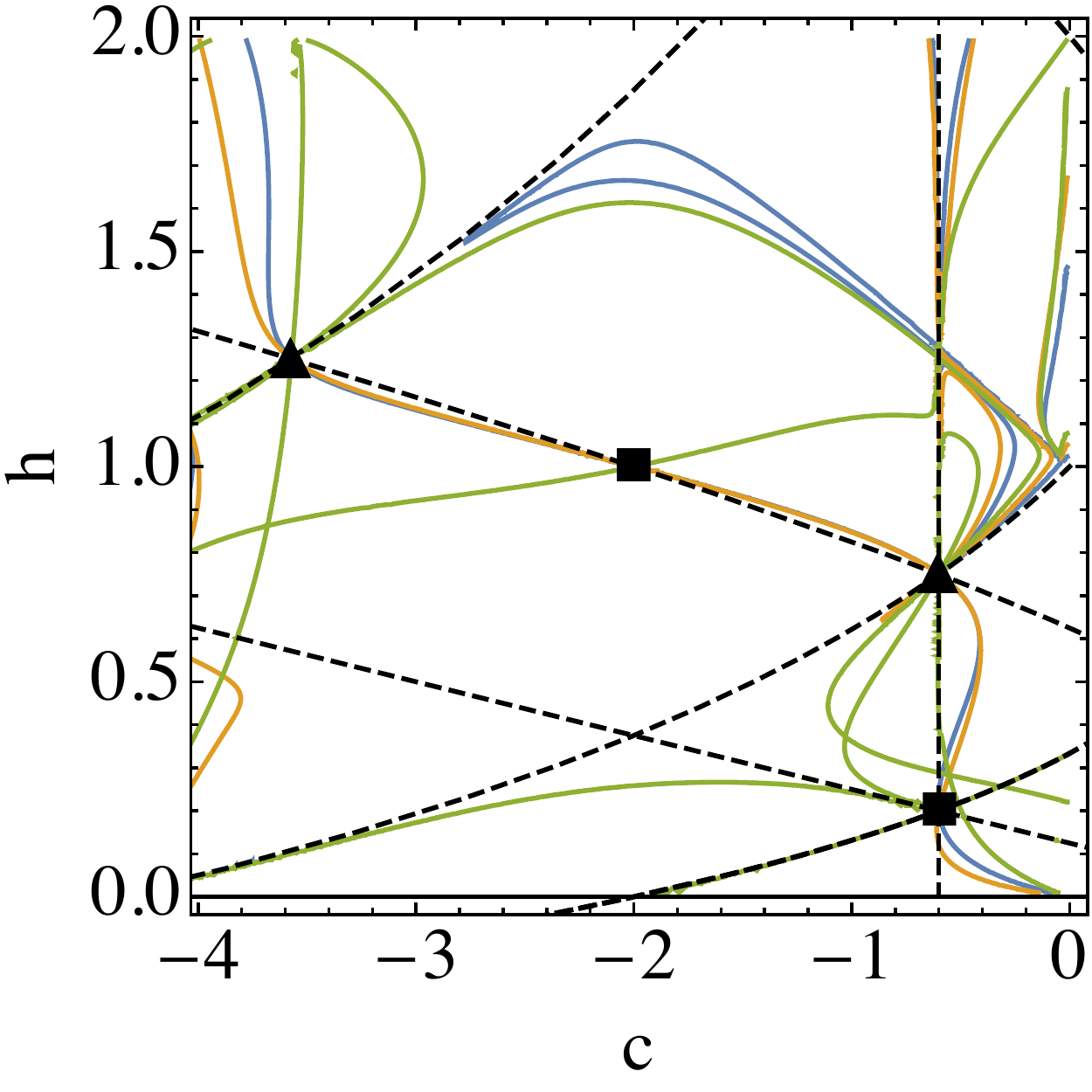}\includegraphics[width=0.47\linewidth]{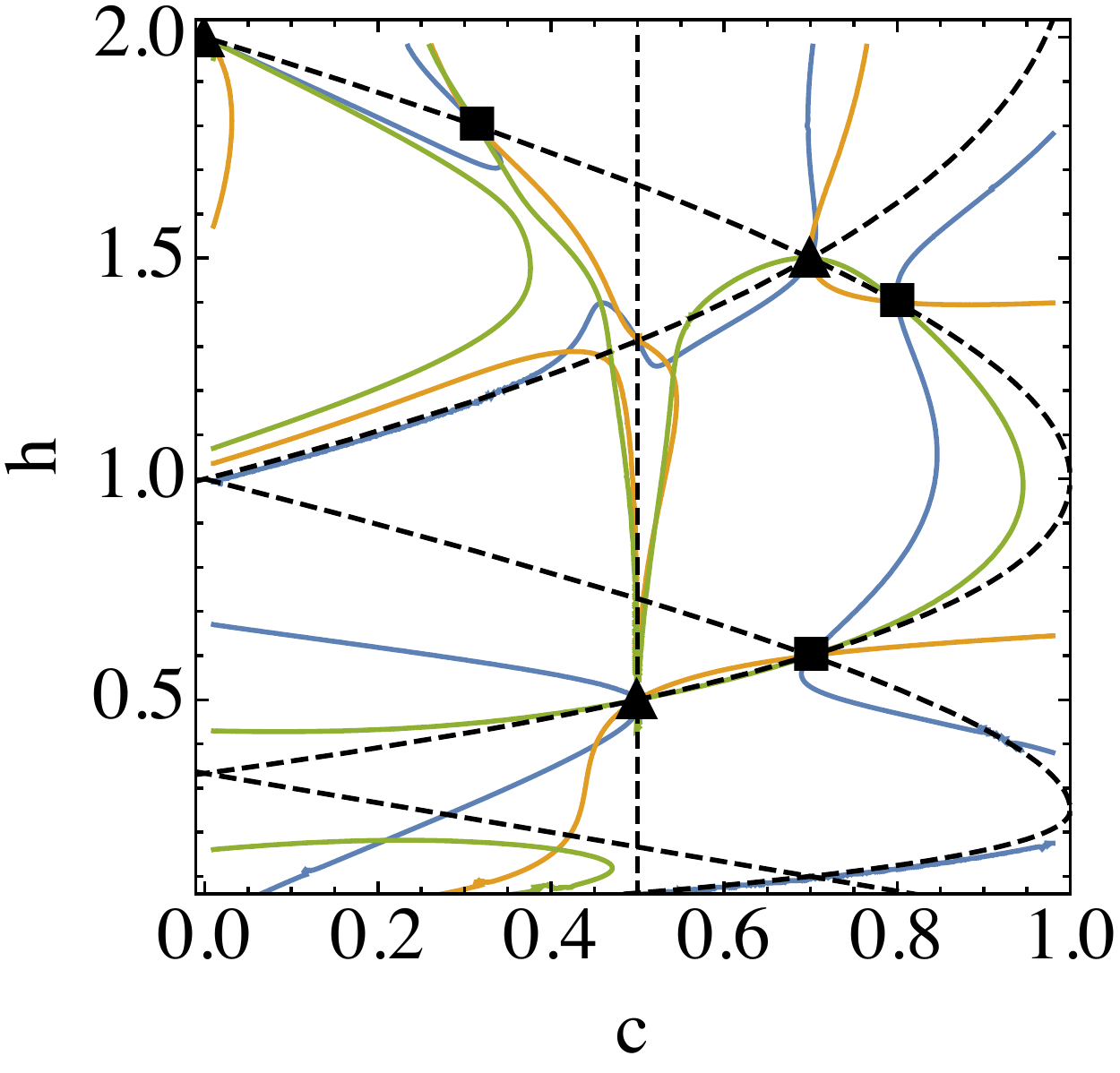}
\end{center}
\caption{Zero contours of the functions \eqref{eq:vir+gli+eqn} in two sub-regions of region $\mathcal R$. Squares are the first few minimal models with OPE closure $[\phi] \times [\phi] = [1]+[\phi]$. Triangles are minimal models with OPE closure $[\phi]\times[\phi]=[1]$. Dashed, black lines are null curves passing through these minimal model values. Points at which all contours intersect are putative solutions to the crossing equation \eqref{eq:vir+gli+eqn}. The only intersections we find correspond precisely to minimal models, plus the $h=1, c=-2$ point.  As explained in the text, the logarithmic CFT at $h=1, c=-2$ looks effectively to the numerics like a $[\phi]\times[\phi] = [1]+[\phi]$ operator algebra.}
\label{fig:ope0c1}
\end{figure}

We now take $N = 2$ and ask whether the crossing condition \eqref{eq:vir+gli+eqn} admits any solution other than those specified by the minimal models in  eq.~\eqref{eq:opetrunc01}. Here we again specialize to the region \eqref{eq:region} and now take $M = 16$ derivatives of the crossing equation. 
Subsets of vanishing minors of the matrix $g^{(m,n)}_{h,\bar h}$ are shown in figure \ref{fig:ope0c1}. 
Points where all minors intersect are putative solutions  to \eqref{eq:vir+gli+eqn} with $N=2$. One of our primary goals was to find new solutions to the crossing equation that are not minimal models, or else to see that all solutions to the truncated OPE ansatz (\ref{eq:ansatz1}) are minimal models themselves.  To make this comparison, we have also plotted in figure \ref{fig:ope0c1} 
the weight $h$ and central charge $c$ of all minimal model operators that satisfy (\ref{eq:ansatz1}) in the region of parameter space shown.  In this region of parameter space,  we find there are \textit{no} solutions to \eqref{eq:vir+gli+eqn} with the given set of derivatives \textit{except} at the minimal models values specified by (\ref{eq:opetrunc01}), and an additional point ($h=1,c=-2$) explained below.  
 Besides this set of minimal models, the solutions plotted in figure \ref{fig:vac_global} also find minimal model cases with  the stronger truncation (\ref{eq:ansatz0}), $[\phi]\times[\phi] = [1]$. This is to be expected, since such an OPE is just a special case of \eqref{eq:ansatz1} where the OPE coefficient of operator $\phi$ vanishes.  

The point $h=1,c=-2$ is a logarithmic CFT that was analyzed in detail in \cite{Gaberdiel:1996kx}.  Technically, in this model the operator $\phi$ is the degenerate operator $(r,s) = (2,1)$, and its fusion  $[\phi]\times[\phi]$ produces the identity and an $(r,s)=(3,1)$ operator (with weight $h_{3,1}=3$), and so it should fall in the class (\ref{eq:ansatz2}) rather than (\ref{eq:ansatz1}).  However, numerically it looks indistinguishable from the fusion rule (\ref{eq:ansatz1}).  The reason for this is that the Virasoro conformal block for $\phi$ itself has a divergent contribution, proportional to the $(3,1)$ conformal block.  To see this explicitly, one can take the conformal block ${\cal F}(c,h_p,h,z)$ at $c=-2, h=1$ but as a general function of the internal operator weight $h_p$, and take the limit $h_p \rightarrow 1$, with the result
\be
{\cal F}(-2, h_p, 1, z) &=& \frac{1}{h_p-1} {\cal F}(-2,3,1,z) + \textrm{reg},
\ee
where ``reg'' denotes terms that are finite at $h_p=1$.  Consequently, in searching for solutions to the bootstrap equation, the algorithm automatically finds OPE coefficients-squared that are $\CO(h_p-1)$ near $h_p \sim 1$. Therefore, the product of the OPE coefficients-squared and the $[\phi]$ block is finite, the only surviving contribution being the $(3,1)$ part of the $[\phi]$ conformal block.

\subsection{Global Block Analysis}

In this section we repeat the analysis working with global, as opposed to Virasoro, conformal primaries, though we will continue to implement a weaker implication of the Virasoro algebra and the truncation (\ref{eq:ansatz1}).  Specifically, we will demand that the scaling dimensions of all operators be either an integer or else $\Delta_\phi$ plus an integer, but we will not impose any relation among the OPE coefficients of different quasi-primaries.\footnote{We will actually demand a somewhat stronger condition that would follow from considering the Virasoro conformal block of pairwise identical scalars, namely that the allowed global blocks have conformal weights $(h,\bar{h})$ that are {\it both} equal to even integers or $\frac{\Delta_\phi}{2}$ plus even integers. See e.g. \cite{Perlmutter:2015iya}. } In this case the central charge can be eliminated at the cost of using an infinite number of global conformal blocks. In 2D, however, the global blocks are simple enough that one can include a large number of global primaries at low computational cost. Moreover, since the OPE is convergent \cite{Pappadopulo:2012jk,Mack:1976pa}, one expects such truncations to give reliable results. This analysis will be exactly like that initiated in \cite{Gliozzi:2013ysa}.

 In this section we change notation slightly. Write the four point function as
	\begin{align}
	\langle \phi(x_1)\phi(x_2)\phi(x_3)\phi(x_4)\rangle &= \frac{g(u,v)}{|x_{12}|^{2\Delta_\phi}|x_{34}|^{2\Delta_\phi}}, \notag \\
	u = \frac{x_{12}^2 x_{34}^2}{x_{13}^2 x_{24}^2}, &\qquad v = \frac{x_{14}^2 x_{23}^2}{x_{13}^2 x_{24}^2},
	\end{align}
where  $u$ and $v$ are related to $z,\bar{z}$ by $u=z \bar{z}, v=(1-z)(1-\bar{z})$.  The function $g(u,v)$ can be expanded in terms of global conformal blocks $G_{\Delta,L}(u,v)$:
	\be
	g(u,v) = \sum_{\Delta, L} p_{\Delta, L} G_{\Delta, L}(u,v),
	\ee
where $p_{\Delta, L} = \lambda^2_{\phi\phi\mathcal O}$ and $\lambda_{\phi\phi\mathcal O}$ is the OPE coefficient with operator
$\mathcal O$ of dimension $\Delta$ and spin $L$ and the sum is over all global primaries appearing the $\phi \times \phi$ OPE. In these variables the crossing condition 
in terms of global blocks reads
	\be
	\sum_{\Delta,L} p_{\Delta,L} \left[v^{\Delta_\phi}G_{\Delta,L}(u,v) - u^{\Delta_\phi}G_{\Delta,L}(v,u) \right] = 0. 
	\label{eq:global+cr+eq}
	\ee
This sum necessarily contains an infinite number of global primaries \cite{Rattazzi:2008pe}. For numerical study the sum must be truncated. This truncation introduces uncontrolled uncertainty in final results which one generally hopes to decrease by including a large number of global primaries. 
	
Truncating the sum with $N$ operators and expanding about the crossing symmetric point $z = \bar z  = 1/2$, one obtains the matrix equation
	\be
	\sum_{\Delta, L}p_{\Delta,L}f^{(m,n)}_{\Delta_\phi,\Delta,L} = 0,  \qquad (m>n\geq 0, ~ m+n\text{ odd}),
	\label{eq:glob+gli+eqn}
	\ee
where
	\be
	f^{(m,n)}_{\Delta_\phi,\Delta,L} = \partial_z^m \partial_{\bar z}^n \left\{[(1-z)(1-\bar z)]^{\Delta_\phi} G_{\Delta,L}(z,\bar z) - (z \bar z)^{\Delta_\phi} G_{\Delta,L}(1-z,1-\bar z) \right\}.
	\ee
As discussed in the previous section, for an OPE including $N$ global primaries, taking $M\geq N$ derivatives of \eqref{eq:glob+gli+eqn} gives a set of $\kappa=\begin{pmatrix}M \\ N \end{pmatrix}$ equations which has a nontrivial solution if and only if all the minors of order $N$ of the matrix $f^{(m,n)}_{\Delta_\phi,\Delta,L}$ are nonvanishing. However, unlike in the previous section, truncating the OPE with $N$ operators is an approximation.

To study the OPE \eqref{eq:ansatz1} by this method we decompose Virasoro primaries $1$  and $\phi$ in terms of global primaries. These are the operators that will appear in \eqref{eq:glob+gli+eqn}. In 2D, these global pimaries will be Virasoro descendants of operators $1$ and $\phi$, and hence their conformal dimensions are \textit{fixed} in terms of the dimension of $\phi$. Using global conformal blocks we therefore only have one free parameter---$\Delta_\phi$. 

\begin{figure}[t!]
\begin{center}
\includegraphics[width=4.0in]{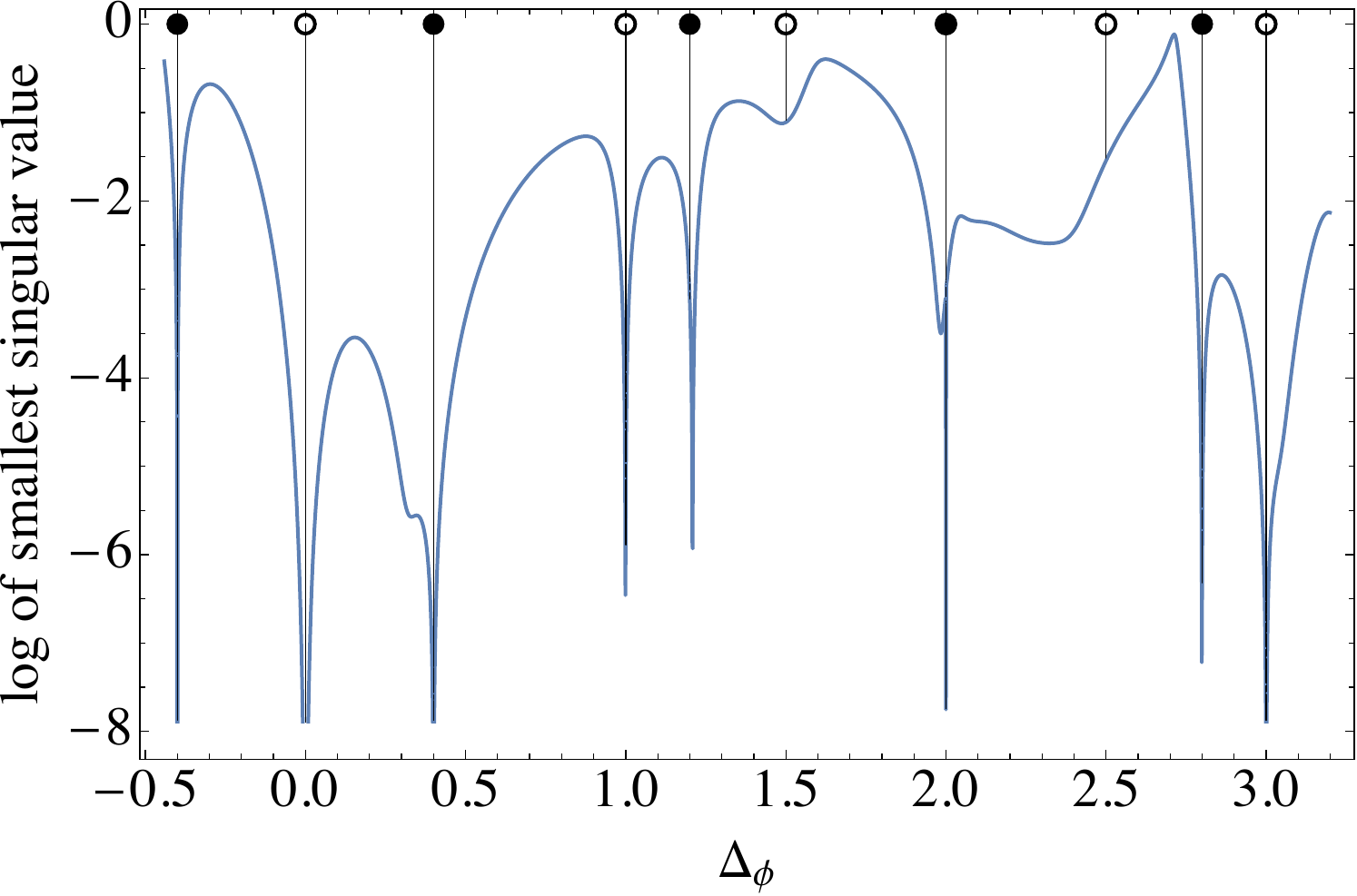}
\end{center}
\caption{The (log of the) smallest singular value of matrix $f^{(m,n)}_{\Delta_\phi,\Delta,L}$. Sharp dips, where this singular value vanishes, correspond to solutions to \eqref{eq:glob+gli+eqn}. Filled circles are minimal models with OPE closure $[\phi] \times [\phi] = [1] + [\phi] $, along with the $(h=1,c=-2)$ log CFT. Open circles are minimal models with OPE closure $[\phi] \times [\phi] = [1]$. 
In this plot we take  $N =111, ~M =112 $. }
\label{fig:single_op_global}
\end{figure}

Rather than looking for vanishing minors of the matrix $f^{(m,n)}_{\Delta_\phi,\Delta,L}$,  we found it easier to look at its singular value decomposition and ask where one of its singular values vanishes. Some reasons for using this approach are discussed at the end of this section. The results can be found in  figure \ref{fig:single_op_global}, in the range $-0.5<\Delta_\phi<3.0$. We see the only dips agree with the Virasoro analysis.  Again, this analysis finds (some) solutions with $[\phi] \times [\phi] = [1]$. Actually, it is a bit surprising that the method is sensitive to the latter set of solutions, since in this case there are roughly half the number of operators in the global block decomposition (the entire $\phi$ module is decoupled) and so there are many more derivatives than operators in matrix $f^{(m,n)}_{\Delta_\phi,\Delta,L}$. The higher derivatives are sensitive to the `missing' operators and so one may have expected this analysis to \textit{not} find this type of OPE at all. 

\begin{figure}[t!]
\begin{center}
\includegraphics[width=4.0in]{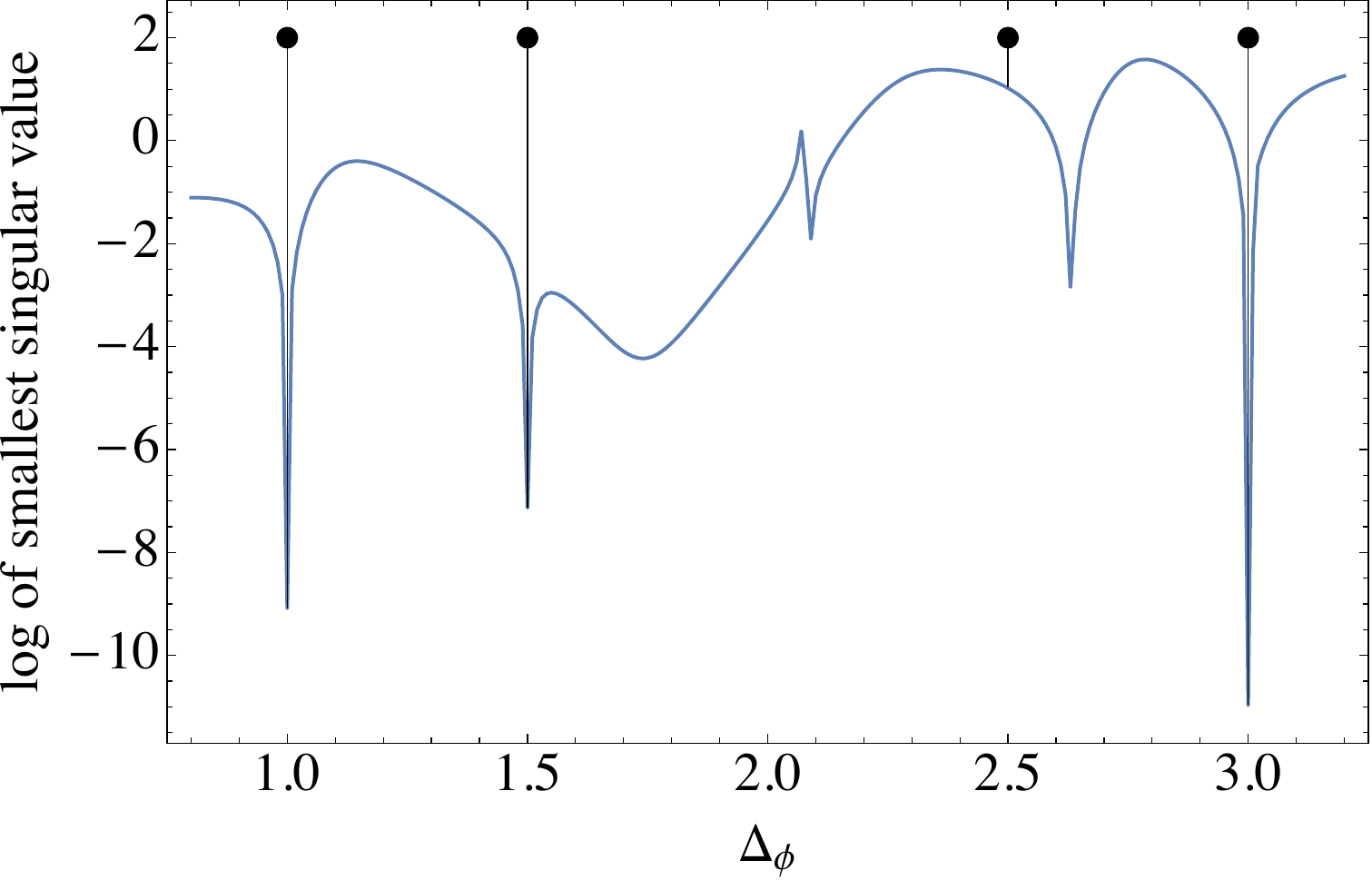}
\end{center}
\caption{The (log of the) smallest singular value of the matrix $f^{(m,n)}_{\Delta_\phi,\Delta,L}$ for the OPE $[\phi]\times[\phi] = [1]$. Filled circles are minimal models with OPE closure $[\phi] \times [\phi] = [1]$. In this case the minimal model with $\Delta_\phi = 3/2$ is clearly found while the model with $\Delta_\phi = 5/2$ is found to within $\sim 10 \%$. In this plot we take $ N = M = 36$. The reason we include a smaller number of operators than in figure \ref{fig:single_op_global} is that we find the matrix $f^{(m,n)}_{\Delta_\phi,\Delta,L}$ becomes numerically unstable more quickly as more operators are included than with the OPE  $[\phi] \times [\phi] = [1] + [\phi]$.}
\label{fig:vac_global}
\end{figure}

It is interesting that the theories the method does find, with $\Delta_\phi = 1$ and 3, are unitary minimal models while those that it misses, $\Delta_\phi = 3/2$ and 5/2, are non-unitary. One possible explanation is that the OPE may converge more rapidly in the unitary case and so the number of global primaries included is sufficient to pick out these theories. As a check we repeat the global block analysis with the OPE $[\phi] \times [\phi] = [1]$. The results are shown in figure \ref{fig:vac_global}. In this case the minimal model with $\Delta_\phi = 3/2$ is found with negligible error while the minimal model with $\Delta_\phi =5/2$ is found within $\sim 10 \%$. \footnote{The $\Delta_\phi=5/2$ point is non-unitary, with $c=-\frac{25}{7}$ and $(r,s)=(1,6)\cong (2,1)$. The numeric situation with the $\Delta_\phi=3$ point is actually somewhat subtle as well.   There is both a unitary and a non-unitary minimal model at $\Delta_\phi=3$, and while the convergence to the correct value of $\Delta_\phi$ is very rapid at this point, the convergence to the correct space of solutions to the OPE coefficients appears to be extremely poor.  It would be interesting to understand the systematics of this issue in more detail.}  


\subsubsection{$[\phi] \times [\phi] = [1]+[\epsilon]$}
\label{sec:one-plus-eps}

\begin{figure}[t!]
\begin{center}
\includegraphics[width=4.0in]{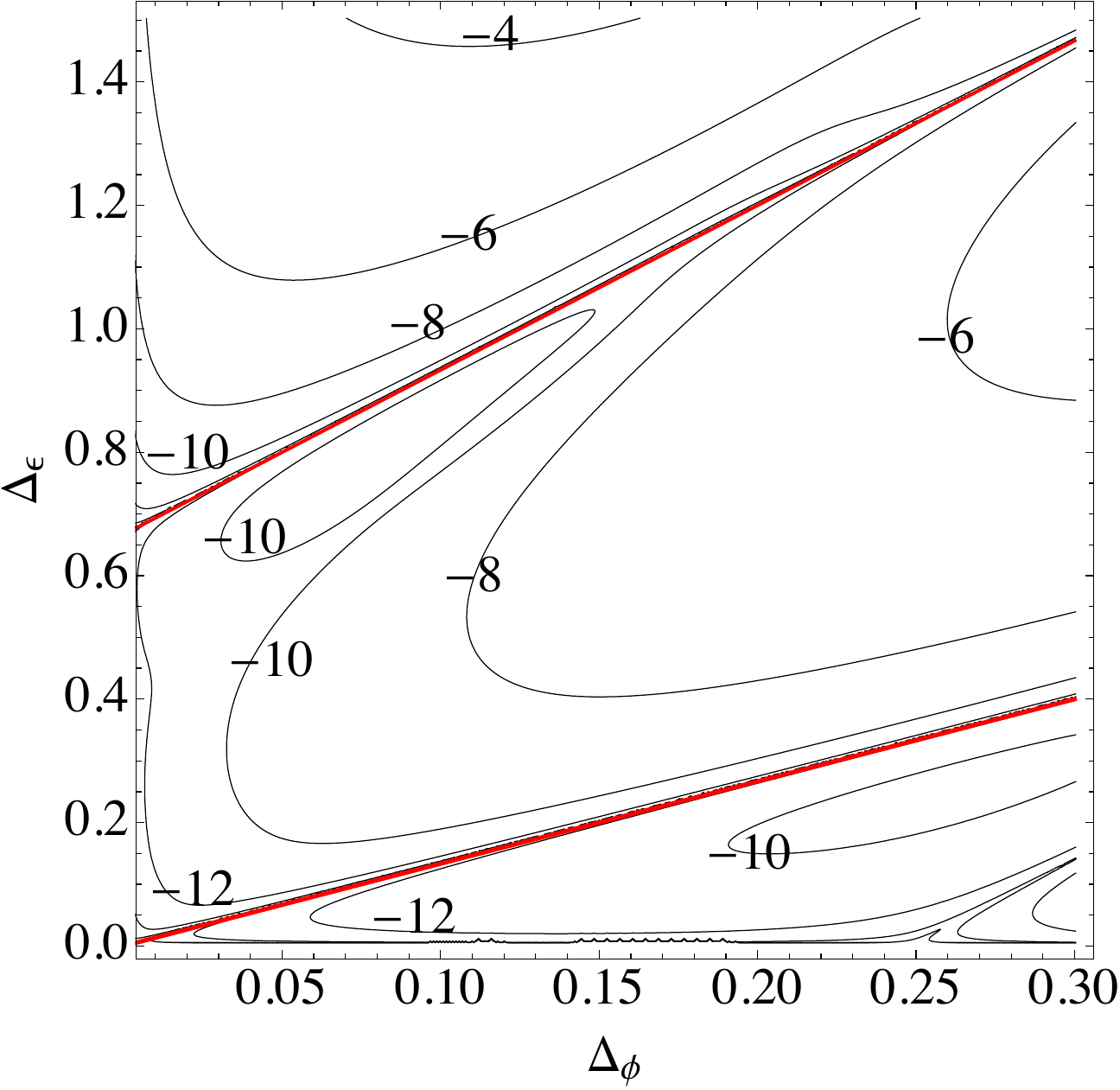}
\end{center}
\caption{Contour plot of the (log of) the smallest singular value of the matrix $f^{(m,n)}_{\Delta_\phi,\Delta,L}$, with OPE closure $[\phi]\times[\phi] = [1]+[\epsilon]$. There is a clear dip into a valley along the red lines. This top line is $\Delta_\epsilon = 8/3 \Delta_\phi + 2/3$ and is well known from \cite{Liendo:2012hy}; it corresponds to operator $\phi$ having a null descendant at level two. The lower red line is $\Delta_\epsilon=4/3\Delta_\phi$ and its origin is discussed in the text.  }
\label{fig:2opMinSV}
\end{figure}

It is not hard to modify our numerics to consider the slightly less trivial OPE
    \be
    [\phi]\times[\phi] = [1]+[\epsilon],
    \ee
for Virasoro primaries $\phi$ and $\epsilon$. 
We do the analysis using global conformal primaries, as explained in the previous section, to avoid explicit reference to central charge.\footnote{This is similar in spirit to the analysis in \cite{ElShowk:2012hu} in that less information is used than is available from the full Virasoro symmetry.  In their case, they impose only the global conformal symmetry and also impose positivity.  In our case, we do not impose positivity, but we impose both global conformal symmetry and the constraint on the spectrum that all operators must have dimension $h_\phi+\textrm{even integers}$ or $h_\epsilon+\textrm{even integers}$, where $h_\phi$ and $h_\epsilon$ are parameters determined by the analysis.} We decompose Virasoro primaries $1$ and $\epsilon$ in terms of global primaries and study the smallest singular value of the matrix $f^{(m,n)}_{\Delta_\phi, \Delta, L}$. All global primaries will be Virasoro descedants of operators $1$ and $\epsilon$ and therefore have conformal dimensions fixed in terms of the dimension of $\epsilon$. Thus we have two free parameters to scan over---$\Delta_\phi$ and $\Delta_\epsilon$.

Contours of the smallest singular value are shown in figure
\ref{fig:2opMinSV}. The two red lines in the figure correspond to
sharp dips, where we find one-parameter families of crossing symmetric
four point functions. The top line is well known, see
e.g. \cite{Liendo:2012hy}, and corresponds to operator $\phi$ having a null
descendant at level two. The equation of this line is \be
\Delta_\epsilon = \frac 83 \Delta_\phi + \frac 23.
	 \label{eq:LeoLine}
	\ee
The lower line is given by
	\be
	\Delta_\epsilon = \frac 43 \Delta_\phi.
	\label{eq:BottomLine}
	\ee

This lower line can also be obtained analytically. To determine its origin, we observe that the identity block in fact decouples; that is, the fusion rule along this line is actually
\begin{equation}
  [\phi] \times [\phi] = [\epsilon]\, .
  \label{eq:BottomLine-fusion}
\end{equation}
To see this, we checked that the the OPE coefficients for the identity
and its descendants vanished (with the normalization $C_{\phi\phi
  \epsilon} = 1$). Furthermore, the central charge along the line can
be determined by comparing the OPE coefficients in the $\epsilon$
block and leads to the relation
\begin{equation}
  \label{eq:BottomLine-central-charge}
  c = 16 \Delta_\phi + 1\, .
\end{equation}
This relation between dimension and central charge is exactly
reproduced by taking $\phi$ to be a `degenerate' operator $\phi_{r,s}$
with Kac indices continued to $r=s=\frac 12$. In fact, the exact four
point function can be obtained by Coulomb gas techniques (reviewed in
appendix \ref{sec:coulomb-gas}): the vertex operator $V_{\frac 12,
  \frac 12} = e^{i \alpha_{\frac 12, \frac 12} \phi}$ has charge
$\alpha_{\frac 12, \frac 12} = \frac{\alpha_0}{2}$ and therefore the
correlator $\langle V_{\frac 12, \frac 12} V_{\frac 12, \frac 12}
V_{\frac 12, \frac 12} V_{\frac 12, \frac 12} \rangle$ trivially
satisfies the neutrality condition $\sum_i \alpha_i = 2\alpha_0$ for
all $\alpha_0$. Hence no screening charge insertions are needed and
the four point function can be immediately written down
\begin{equation}
  \left\langle V_{\frac 12,
      \frac 12}(\infty) V_{\frac 12,
      \frac 12}(1) V_{\frac 12,
      \frac 12}(z, \bar z) V_{\frac 12,
      \frac 12}(0) \right \rangle = \vert z(1-z) \vert^{\alpha_0^2} = \vert z (1 - z) \vert^{{-}2\Delta_\phi/3}\, ,
\end{equation}
where we've used the dimension of $V_{\frac 12, \frac 12}$,
$\Delta_{\frac 12, \frac 12} = {-} \frac{3 \alpha_0^2}{2}$. The
parameter $\alpha_0$ fixes the central charge to $c = 1- 24
\alpha_0^2$, and so we see that $c = 16 \Delta_\phi + 1$. Note that
$\Delta_\phi > 0$ implies $\alpha_0$ is purely imaginary and therefore
$c >1$. This four point function is manifestly crossing symmetric and
has a unique exchanged operator, whose dimension we can easily extract
and which obeys the advertised relation \eqref{eq:BottomLine}. We note
that this exchanged operator can also be written as a `degenerate'
operator, with $(r,s) = (0,0)$.

Finally, we note that generalized free fields (GFFs) and free scalar theories do not appear as solutions in the above plot.  This is because they do not satisfy the condition we imposed on the spectrum as a weak consequence of Virasoro symmetry, namely that all global primary operators have weights $(h,\bar{h})$ that are either even integers or $\frac{\Delta_\epsilon}{2}$ plus even integers.

\subsubsection{Singular Values vs Minors} 

\begin{figure}[t!]
\begin{center}
\includegraphics[width=0.45\linewidth]{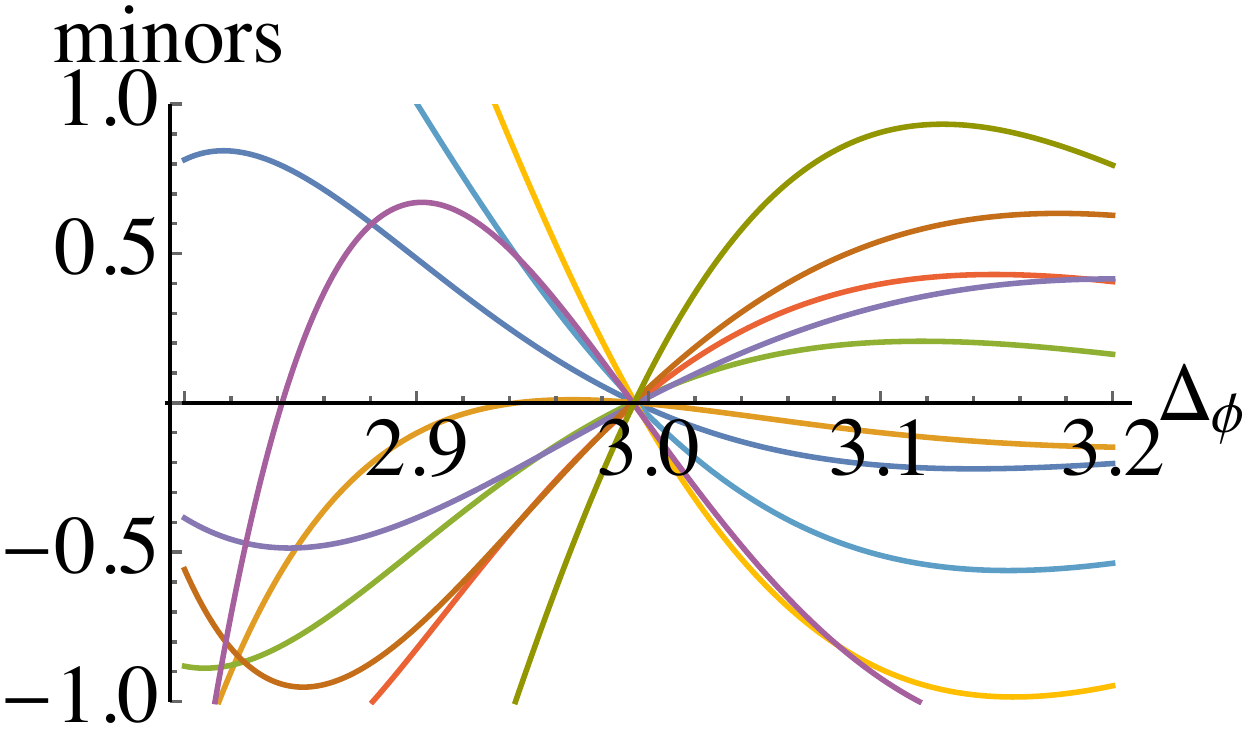}\includegraphics[width=0.47\linewidth]{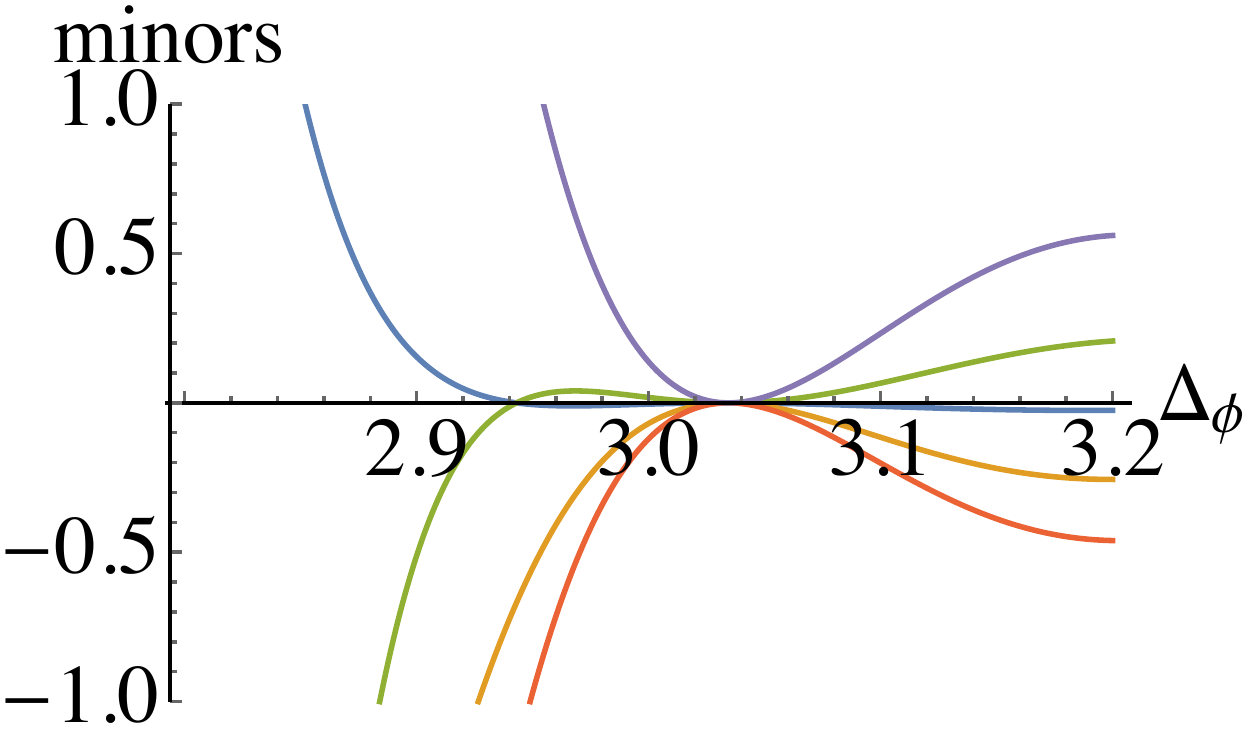}
\end{center}
\caption{The first panel shows minors with roots spread around the known value $\Delta_\phi=3$. The second panel shows different minors with simultaneous roots at a value off from $\Delta_\phi=3$ by $\sim 1 \%$.}
\label{fig:minors}
\end{figure}
\begin{figure}[t!]
\begin{center}
\includegraphics[width=0.45\linewidth]{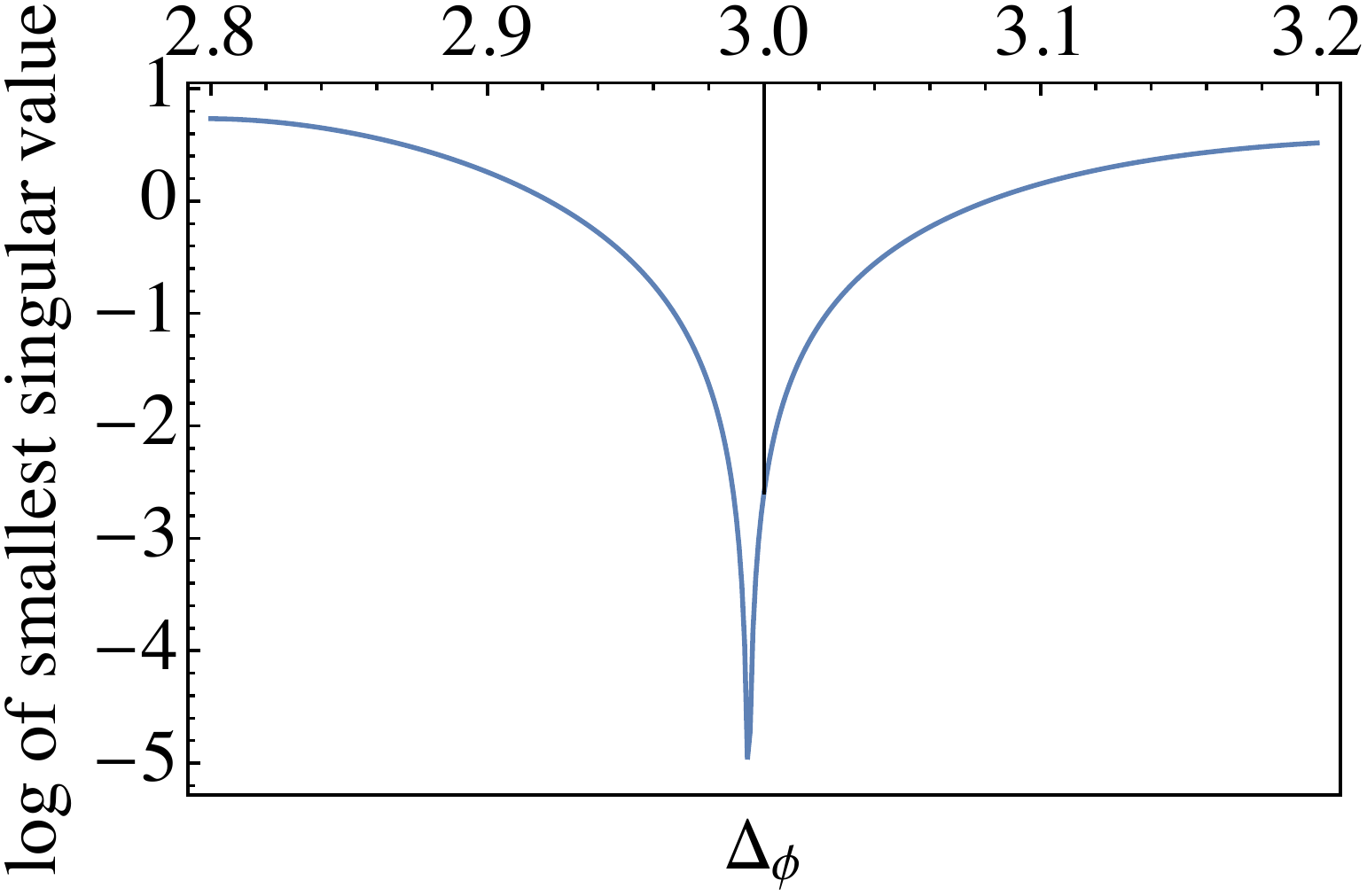}
\end{center}
\caption{A plot of the smallest singular value of the matrix $\mathsf f$. The number of operators is the same as in figure \ref{fig:minors}. In this case there is a sharply defined minimum at $\Delta_\phi \approx 2.994$. }
\label{fig:SVcompare}
\end{figure}

In this subsection we note some of the benefits of studying singular values as opposed to minors. Firstly, optimization problems are far more robust than root finding. Therefore it will in general be more efficient to minimize the smallest singular value of the matrix $f^{(m,n)}_{\Delta_\phi,\Delta,L}$ over the space of unknown dimensions than it would be to find simultaneous roots of its minors. This will be especially important if one wants to pursue this program in $D>2$, including many operators. In this work we were able to include a large number of operators because the dimensions of Virasoro descendants were fixed, requiring us to scan over only one or two operator dimensions. In higher $D$, where one no longer has the constraints of Virasoro symmetry, all operator dimensions must be left variable. 

The second point has to do with how the roots of minors of $f^{(m,n)}_{\Delta_\phi,\Delta,L}$ organize themselves. For a low number of operators, one might expect a scatter of roots around the right solution. In practice, however, one often finds  multiple points where many (but not all) minors have simultaneous roots. This is illustrated in figure \ref{fig:minors} in the case of the  tricritical Ising model, for an operator $\phi$ which has OPE closure $[\phi]\times[\phi]=[1]$ and dimension $\Delta_\phi=3$. The first panel shows a collection of minors with roots spread around the exact value. The second panel shows a different collection of minors with simultaneous roots at a value of $\Delta_\phi$ off from the correct value by $\sim 1 \%$. This discrepancy is not very significant, and one can check such artifacts disappear as more operators are included, but it would be nice to have a method that is more robust. In this regard we found singular values more useful. For the same number of operators, a plot of the smallest singular value of the matrix $f^{(m,n)}_{\Delta_\phi,\Delta,L}$ is shown in figure \ref{fig:SVcompare}. In this case there is a single pronounced dip near the exact value and doing the minimization yields $\Delta_\phi \approx 2.994$---an error of $\sim 0.2 \%$.

\section{Crossing Matrix Analysis}
\label{sec:tesch}

In this section, we sketch an argument to demonstrate that the finite
operator algebras studied in previous sections imply the operators
must be degenerate. To do so, we borrow some technology from
\cite{Teschner1,Teschner2,Teschner3} derived in the context of
Liouville theory. In particular, we will use an explicit integral
expression for the holomorphic crossing matrix to argue that the fusion
rule $[\phi] \times [\phi] = [1] + [\phi]$ implies that $\phi$ is a
degenerate operator. Throughout we'll restrict entirely to the
holomorphic sector, as we are only concerned in the structure of the
holomorphic crossing matrix.

The basic setup, reviewed for example in \cite{TeschnerLec,Teschner1},
is analogous to the Coulomb gas description of the minimal models
described in the appendices. Namely we construct representations of
the Virasoro algebra with central charge $c=1+6Q^2$ out of a (chiral)
free scalar $\phi$.\footnote{The Liouville $Q$ parameter (which has a
  slightly different normalization compared to the discussion in the
  appendix) is related to the Coulomb gas parameter $\alpha_0$ by $Q=
  2i\alpha_0$.} In this approach to quantizing Liouville theory, the
Hilbert space factorizes as a direct sum over a continuum of free
scalar Fock spaces with different `momenta' $p$, ${\cal H} \sim \int d
p\, {\cal F}_p$, and the primaries in the theory are constructed as
screened exponentials $V_\alpha(z)$ of the scalar; here ${\cal F}_p$
is a highest weight Virasoro representation space with $c=1+6Q^2$ and
$\Delta = p^2+\frac{Q^2}{4}$, generated by the screened exponential
with momentum $p$. We will primarily label the exponentials $V_\alpha$
by their charge $\alpha$, which is related to the momentum $p$ and
dimension $\Delta$ by
\begin{equation}
  \alpha = \frac{Q}{2} + i p\, ,  \qquad \Delta = p^2 + \frac{Q^2}{4} = \alpha (Q - \alpha)\, .
\end{equation}
Finally, we note that $Q$ is related to the Liouville parameter $b$
via $Q = b + b^{{-}1}.$

The primary object of interest to us will be the holomorphic crossing
matrix $F_{\alpha_s,\alpha_t} \left[\begin{smallmatrix} \alpha_2 &
    \alpha_3 \\ \alpha_1 & \alpha_4
  \end{smallmatrix}\right]$. It is defined as usual by relating the
conformal block decompositions in the $s$-and $t$-channels, as shown
in \eqref{eq:holomorphic-fusion-matrix}; however, due to the
continuous spectrum, the `matrix' is actually an integral kernel. That
is, when expanding an $s$-channel block in the $t$-channel, we will
generically obtain a continuum of conformal blocks. Our goal is to
understand when this continuum can be restricted to a discrete sum of
conformal blocks. Therefore we need to analyze the crossing kernel and
determine when it can be written as a linear combination of $\delta$
functions.

\subsection{Overview}

It was shown in \cite{Teschner3} that the crossing kernel can be written in the form
\be
\label{eq:Fmat-schem}
F_{\alpha_s, \alpha_t}
\begin{bmatrix} \displaystyle
  \alpha_2 & \alpha_3 \\ \alpha_1 & \alpha_4
\end{bmatrix} &=& P(\alpha_i) \int_{\cal C} \mathrm{d}u \,{\cal I}(u; \alpha_i)
\ee where $P(\alpha_i)$ is a function of the $\alpha$'s (both for the
external and exchanged operators), and the integrand ${\cal
  I}(u; \alpha_i)$ is a function of these $\alpha$'s as well as the
integration variable $u$, which is integrated over some contour ${\cal
  C}$ plotted in figure \ref{fig:poles-and-contour}.  We review the
definition of $P(\alpha_i)$, ${\cal I}(u; \alpha_i)$ and ${\cal C}$ in
detail below, but first we will discuss the essential points of their
qualitative behavior.
  
  As mentioned above, if we take any single $s$-channel block and
  expand it in the $t$-channel, then generically we obtain a
  continuum of blocks; to avoid such continuous operator algebras, the
  kernel must localize to a discrete set of points determined by the
  $\alpha_i$, $\alpha_s$, and $c$. For simplicity, we'll focus
  entirely on the case where the external operators are
  identical scalars, $\alpha_i = \alpha$, with the fusion rule $[\phi]
  \times [\phi] = [1] + [\phi]$. Our strategy will be to fix $\alpha_s
  = 0$ (more precisely, we take $\alpha_s = \varepsilon$ and consider
  the behavior as $\varepsilon \to 0$) and search for the situations
  in which \eqref{eq:Fmat-schem} is singular at particular values of
  $\alpha$, $\alpha_t$ while vanishing elsewhere.

\begin{figure}
    \centering
    \begin{tikzpicture}[scale=5]
      \draw[green,dashed,thick] (0,0) .. controls (.5, .05) and
      (.5,.9) .. (1,1) node[at start, anchor=south] {${\cal C}$};

      \node[red] at (.25,.5) {$\times$}; \node[anchor=north,red] at
      (.25,.5) {$u_a$};

      \node[blue] at (.75,.5) {$\times$};
      \node[anchor=north,blue] at (.75,.5) {$u_b(\alpha,\alpha_t)$};

      \draw (.05,.95) -- (.05, .85) -- (.15,.85);
      \node[anchor=south west] at (.05,.85) {$u$};
  

      \draw[double distance=1pt,-stealth,thick] (1,.5) -- (1.5, .5)
      node[midway,anchor=north] {vary $\alpha, \alpha_t$};

      \draw (1.55,.95) -- (1.55, .85) -- (1.65,.85);
      \node[anchor=south west] at (1.55,.85) {$u$};
      
      \draw[green,dashed,thick] (1.5,0) .. controls (2, .05) and (2,.9)
      .. (2.5,1) node[at start, anchor=south] {${\cal C}$};
  
      \node[red] at (1.75,.5) {$\times$};
      \node[blue] at (1.825,.5) {$\times$};
      
      \draw[green,dashed,thick] (1.825,.5) circle [radius=1pt];

      \draw[-stealth] (1.825,.65) -- (1.825,.55) node[at start, anchor=south]
      {Residue $\propto \frac{1}{{\color{blue} u_b}-{\color{red}u_a}}$};
    \end{tikzpicture}
    \caption{Schematic illustration of singular contributions to the
      integral over $u$ in the crossing kernel. The contour ${\cal C}$,
      depicted in green, is defined by separating the poles from the
      numerator (blue) from the poles in the denominator (red) for a
      particular regime of the parameters $\alpha_i, \alpha_s,
      \alpha_t$. As we vary these parameters, the poles will move
      around in the $u$-plane and the contour must deform
      accordingly. Here, the pole at $u_b(\alpha,\alpha_t)$ moves left
      and collides with $u_a$ as $\alpha$ and $\alpha_t$ are
      changed. As the contour must run between $u_a$ and $u_b$, we
      pick up the residue at $u_a$, which is itself singular when $u_a
      \approx u_b$.}
    \label{fig:poles-collide}
  \end{figure}
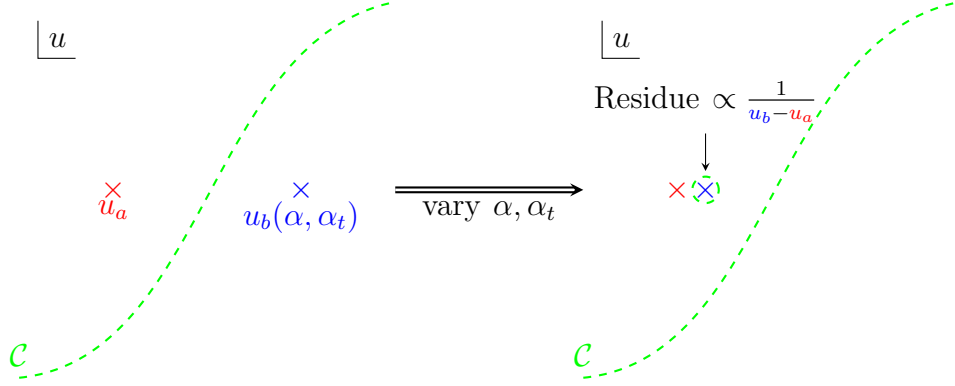

  The singularities can come from either the prefactor or the
  integral. Let's first consider the prefactor $P(\alpha_i)$. As can
  be seen from the explicit form of $P(\alpha_i)$ given below,
  $P(\alpha_i)= 0$ for generic values of $\alpha$, $\alpha_t$ whenever
  exchanging the identity operator in the $s$-channel, $\alpha_s =
  0$. However, for particular values of $\alpha$ and $\alpha_t$,
  namely when they correspond to degenerate operators, the zero in
  this prefactor is cancelled (or at least its order reduced) by
  additional singularities. This suggests that whenever the identity operator is exchanged in
  the $s$-channel, the integral over $u$ must be singular for any
  operators which contribute to the fusion algebra. That is, we can
  ignore any regular part of the integral since the zeros in the
  prefactor will render such terms unimportant. The integrand ${\cal
    I}(u;\alpha)$ is a ratio of complicated meromorphic functions,
  with the pole locations dependent upon the parameters $\alpha$ and
  $\alpha_t$. As we tune these parameters, the poles will move around
  in the $u$-plane, but thanks to analyticity, we are free to deform
  the contour away from the poles. The only way the integral can
  develop singularities is when two (or more) poles pinch the
  integration contour (see figure~\ref{fig:poles-collide}). The
  crucial point is that this pole collision only occurs for particular
  values of $\alpha$, $\alpha_s$, and $\alpha_t$. The singularities
  arising from this pole collision then compete with the zeros from
  the prefactor, producing either a finite value (corresponding
  to a continuous spectrum in the algebra) or a singular contribution
  to the crossing kernel (corresponding to a discrete operator in the
  operator algebra). In the following subsection we will go through this argument in more detail.

\subsection{Detailed Analysis}

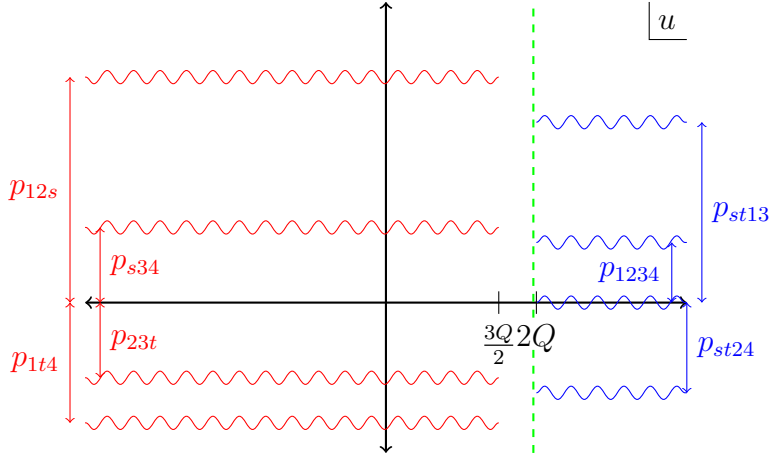
\begin{figure}
  \centering
  \begin{tikzpicture}[scale=2]
    \draw [<->, thick] (-2,0) -- (2,0);
    \draw [<->, thick] (0,-1) -- (0,2);
    
    \draw[red,style={decorate, decoration=snake}] (.75,1.5)
    --(-2,1.5);
    \draw[red,style={decorate, decoration=snake}] (.75,.5)
    --(-2,.5);
    \draw[red,style={decorate, decoration=snake}] (.75,-.5)
    --(-2,-.5);
    \draw[red,style={decorate, decoration=snake}] (.75,-.8)
    --(-2,-.8);

    \draw[blue,style={decorate, decoration=snake}] (1,0) -- (2,0);
    \draw[blue,style={decorate, decoration=snake}] (1,.4) -- (2,.4);
    \draw[blue,style={decorate, decoration=snake}] (1,-.6) -- (2,-.6);
    \draw[blue,style={decorate, decoration=snake}] (1,1.2) -- (2,1.2);

    \draw[red,<->] (-2.1,0) -- (-2.1,1.5) node[midway,anchor=east]
    {$p_{12s}$};
    \draw[red,<->] (-1.9,0) -- (-1.9,.5) node[midway,anchor=west]
    {$p_{s34}$};
    \draw[red,<->] (-1.9,0) -- (-1.9,-.5) node[midway,anchor=west]
    {$p_{23t}$};
    \draw[red,<->] (-2.1,0) -- (-2.1,-.8) node[midway,anchor=east]
    {$p_{1t4}$};

    \draw[blue,<->] (2,0) -- (2,-.6) node[midway,anchor=west] {$p_{st24}$};
    \draw[blue,<->] (1.9,0) -- (1.9,.4) node[midway,anchor=east] {$p_{1234}$};
    \draw[blue,<->] (2.1,0) -- (2.1,1.2) node[midway,anchor=west] {$p_{st13}$};


    \draw[green,dashed,thick,rounded corners] (.98,-1) -- (.98,-.3);
    \draw[green,dashed,thick,rounded corners] (.98,-.1) -- (.98,2);

    \draw (1.75,2) |- (2,1.75)
    node[anchor=south east] {$u$}; \draw (.75,.075) -- (.75,-.075)
    node[anchor=north] {$\tfrac{3Q}{2}$}; \draw (1,.075) --
    (1,-.075) node[anchor=north] {${\textstyle 2Q}$};
  \end{tikzpicture}
  \caption{Illustration of the pole structure of the integrand in
    equation \ref{eq:rwsymbdef}. The blue (red) curves denote poles
    arising from the numerator (denominator) and the green dashed line
    denotes the integration contour, which separates the two sets of
    poles. The blue and red curves are shorthand for a lattice of
    simple poles whose real parts differ by $(i b + j b^{-1})$ with
    $i,j \in \mathbb{Z}^{\geq 0}$ (see e.g.~equations
    \eqref{eq:poles-num1}--\eqref{eq:poles-den3}). The structure is
    depicted for $b\in \R$ and $\alpha_i$, $\alpha_s$, $\alpha_t \in
    \frac{Q}{2} + i \R$ (recall $p = \Im \alpha$), and again a
    multi-index indicates summation, e.g.~$p_{12} = p_1 + p_2$. More
    general regimes are obtained via analytic continuation.}
  \label{fig:poles-and-contour}
\end{figure}

As shown in \cite{Teschner3}, building on earlier work
\cite{Teschner1, Teschner2}, the holomorphic crossing matrix takes the
form
\begin{align}
  \label{eq:tesch-fusion}
  F_{\alpha_s, \alpha_t}
  \begin{bmatrix} \displaystyle
    \alpha_2 & \alpha_3 \\ \alpha_1 & \alpha_4
  \end{bmatrix}
  ={}& \mu(\alpha_t) {\cal N}(\alpha_i,\alpha_s,\alpha_t) {\cal
    M}(\alpha_i,\alpha_s,\alpha_t) 
  \begin{Bmatrix}
    \alpha_1 & \alpha_3 & \alpha_s \\ \alpha_2 & \alpha_4 & \alpha_t
  \end{Bmatrix}\, ,
\end{align}
where:
\begin{align}
  {\cal N}(\alpha_i,\alpha_s,\alpha_t) ={}& \frac{N(\alpha_s,
    \alpha_2, \alpha_1) N(\alpha_4, \alpha_3, \alpha_s)}{N(\alpha_t,
    \alpha_3, \alpha_2) N(\alpha_4, \alpha_t, \alpha_1)}\, , \\
  {\cal M}(\alpha_i,\alpha_s,\alpha_t) ={}& \frac{M(\alpha_t,
    \alpha_3, \alpha_2) M(\alpha_4, \alpha_t, \alpha_1)}{M(\alpha_s,
    \alpha_2, \alpha_1) M(\alpha_4, \alpha_3,
    \alpha_s)}\,, \\
  N(\alpha_3, \alpha_2, \alpha_1) ={}& \frac{\Gamma_b(2Q- 2\alpha_3)
    \Gamma_b(2 \alpha_2) \Gamma_b(2\alpha_1) \Gamma_b(\alpha_{12} -
    \alpha_3) }{\Gamma_b(2Q-\alpha_{123}) \Gamma_b(\alpha_{13} -
    \alpha_2) \Gamma_b(\alpha_{23} - \alpha_1)}\, , \label{eq:teschNdef}\\
  M(\alpha_3, \alpha_2, \alpha_1) ={}& \left[ \frac{S_b
      (\alpha_{123}-Q) S_b(\alpha_{12} - \alpha_3)}{S_b(\alpha_{13} -
      \alpha_2) S_b(\alpha_{23} - \alpha_1)}
  \right]^{1/2} \label{eq:teschMdef}  \\
  \mu(\alpha) ={}& \left\vert S_b(2\alpha) \right\vert^2\,
  , \label{eq:teschmudef}
\end{align}
and $\begin{Bmatrix}
    \alpha_1 & \alpha_3 & \alpha_s \\ \alpha_2 & \alpha_4 & \alpha_t
  \end{Bmatrix}$ is the Racah-Wigner coefficient for the quantum group
  $U_q(sl(2,\mathbb{R}))$, given by the following integral:
\begin{equation}
  \label{eq:rwsymbdef}
  \begin{split}
    \begin{Bmatrix}
      \alpha_1 & \alpha_3 & \alpha_s \\ \alpha_2 & \alpha_4 & \alpha_t
    \end{Bmatrix} ={}& \Delta(\alpha_1, \alpha_2, \alpha_s)
    \Delta(\alpha_s, \alpha_3, \alpha_4) \Delta(\alpha_t, \alpha_3,
    \alpha_2) \Delta(\alpha_4, \alpha_t, \alpha_1)
    \times  \\
    &\int_{{\cal C}} \mathrm{d} u \frac{S_b(u-\alpha_{12s})
      S_b(u-\alpha_{s34}) S_b(u-\alpha_{23t})
      S_b(u-\alpha_{1t4})}{S_b(u+Q-\alpha_{1234})
      S_b(u+Q-\alpha_{st13}) S_b(u+Q-\alpha_{st24}) S_b(u-Q)}\,
    . 
  \end{split}
\end{equation}
In \eqref{eq:teschNdef}--\eqref{eq:rwsymbdef}, a multi-indexed
$\alpha$ indicates summing over the corresponding $\alpha_i$,
e.g.~$\alpha_{ij} = \alpha_i + \alpha_j$, and $\Gamma_b$, $S_b$ are
special functions whose pertinent features are reviewed in appendix
\ref{sec:funcs}. The function $\Delta(\alpha_1,\alpha_2, \alpha_3)$ is given by:
\begin{align}
  \Delta(\alpha_1, \alpha_2, \alpha_3) ={}& \left(
    \frac{S_b(\alpha_{123} - Q)}{S_b(\alpha_{12} - \alpha_3)
      S_b(\alpha_{13} - \alpha_2) S_b(\alpha_{23} - \alpha_1)}
  \right)^{1/2}\, . \label{eq:deltadef}
\end{align}
The contour ${\cal C}$ in \eqref{eq:rwsymbdef} is defined by
separating the poles of the numerator from the zeros of the
denominator and approaching $2Q+i \R$ near infinity, as shown in
figure~\ref{fig:poles-and-contour}.

For the case of interest, namely identical external operators and
exchanging the identity operator in the $s$-channel, we take $\alpha_i
= \alpha$ and $\alpha_s = \varepsilon$, with $\varepsilon$ to be sent
to zero at the end of the day. As mentioned above, the prefactor
$P(\alpha_i,\alpha_s,\alpha_t)$ vanishes with $\alpha_s = 0$ and
generic $\alpha$, $\alpha_t$. More explicitly, we see from equations
\eqref{eq:tesch-fusion}--\eqref{eq:teschmudef} that, in this limit,
the prefactor behaves, up to a phase factor, as
\begin{align}
  \label{eq:pref-sid}
  P \propto{}& \frac{1}{S_b(\varepsilon)^2} \mu(\alpha_t) \left(
    \frac{\Gamma_b(2\bar\alpha - \alpha_t)
      \Gamma_b(\alpha_t)}{\Gamma_b(2\bar\alpha)}\right)^2
  \frac{\Gamma_b(2Q)
    \Gamma_b(Q)}{\Gamma_b(2\alpha_t)\Gamma_b(2\bar\alpha_t)}
  \frac{S_b(2\alpha-\bar\alpha_t)^2}{S_b(2\alpha)
    S_b(\alpha_t)^3S_b(2\alpha-\alpha_t)} \, .
\end{align}
Here for brevity we've introduced $\bar \alpha = Q - \alpha$. Thus for
generic $\alpha$ and $\alpha_t$, the prefactor has a double zero as
$\alpha_s =\varepsilon \to 0$ due to the factor of
$S_b(\varepsilon)^2$ (see equation (\ref{eq:sbpoles})).

\begin{figure}
  \centering
  \begin{tikzpicture}[scale=2]

    \draw[double distance=1pt,red,style={decorate, decoration=snake}] (.35,.75)
    --(-2,.75) node[at end, anchor = east] {$u_{\text{num},2}$};
    \draw[double distance=1pt,red,style={decorate, decoration=snake}] (.2,.25)
    --(-2,.25) node[at end, anchor = east] {$u_{\text{num},1}$};

    \draw[blue,style={decorate, decoration=snake}] (1,0) -- (2,0)
    node[at end,anchor = west] {$u_{\text{den},1}$};
    
    \draw[double distance=1pt,blue,style={decorate, decoration=snake}]
    (.5,.75) -- (2,.75) node[at end,anchor = west] {$u_{\text{den},3}$};;
    
    \draw[blue,style={decorate, decoration=snake}] (.8,1) -- (2,1)
    node[at end,anchor = west] {$u_{\text{den},2}$};;
    
    \draw [<->, thick] (-2,0) -- (2,0);
    \draw [<->, thick] (0,-.5) -- (0,1.5);
    
    \draw[|<->|] (.35,.65) -- (.5,.65) node[midway, anchor=north]
    {$\varepsilon$};

    \draw (1.75,1.5) |- (2,1.25) node[anchor=south east] {$u$};

    \draw (1,.075) -- (1,-.075) node[anchor=north] {${\textstyle
        2Q}$};
  \end{tikzpicture}
  \caption{Illustration of the pole structure (equations
    \eqref{eq:poles-num1}--\eqref{eq:poles-den3}) of the integrand
    ${\cal I}$ for $\alpha_i = \alpha$ and $\alpha_s
    =\varepsilon$. Here the doubled curves indicate double poles. As
    $\varepsilon\to 0$, the (double) poles at $u_{\text{num},2}^{0,0}$
    and $u_{\text{den,3}}^{0,0}$ collide. To deal with this, the
    integration contour is pushed through $u_{\text{den,3}}^{0,0}$,
    which introduces a singular contribution to the integral.}
  \label{fig:intgd1-poles}
\end{figure}
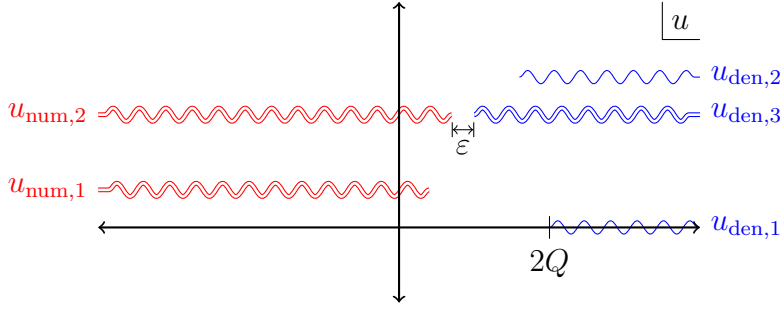

Now we turn to the integral. Plugging in the appropriate values for
the $\alpha$s, the integrand takes the form:
\begin{equation}
  \label{eq:intgd1}
  {\cal I}(u;\alpha,\varepsilon,\alpha_t) =
  \frac{[S_b(u-2\alpha-\varepsilon) S_b(u-2\alpha -
    \alpha_t)]^2}{S_b(u+Q-2\alpha-\alpha_t-\varepsilon)^2 S_b(u-Q)
    S_b(u+Q - 4 \alpha)}\, .
\end{equation}
The singularities of this integrand (arising from the poles and zeros
of $S_b(x)$, given in equation \eqref{eq:sbpoles}) are located at the
following values for $u$:
\begin{align}
  \text{Poles of numerator:} \quad u_{\text{num},1}^{i,j} ={}& 2\alpha
  + \varepsilon - (i b + j b^{-1})\, , \label{eq:poles-num1}\\
  u_{\text{num},2}^{i,j} ={}& 2\alpha +\alpha_t - (i b + j b^{-1})\,
  , \label{eq:poles-num2} \\
  \text{Zeros of denominator:} \quad u_{\text{den},1}^{i,j} ={}& 2Q +i
  b + j b^{-1}\, , \label{eq:poles-den1} \\
  u_{\text{den},2}^{i,j} ={}& 4\alpha +i b + j b^{-1}\, , \label{eq:poles-den2} \\
  u_{\text{den},3}^{i,j} ={}& 2\alpha+ \alpha_t + \varepsilon +i b + j
  b^{-1} \label{eq:poles-den3}\, .
\end{align}
Here $i,j$ are non-negative integers, and we note that all of the
poles from the numerator as well as the poles at
$u_{\text{den},3}^{i,j}$ are double poles. This pole structure is
depicted in figure~\ref{fig:intgd1-poles}.

Note that the (double) poles at $u_{\text{num},2}^{0,0}$ and
$u_{\text{den},3}^{0,0}$ overlap as $\varepsilon \to 0$. Since the
integration contour separates these poles, this collision introduces a
singularity of the type discussed above. As illustrated in
figure~\ref{fig:poles-collide}, we push the contour through
$u_{\text{den},3}^{0,0}$, picking up its residue in the process and
yielding
\begin{align}
  \int_{\cal C}\mathrm{d}u\, {\cal I} \propto{}& \mathrm{Res}\left({\cal I},
    u_{\text{den},3}^{0,0}\right) + \left(\text{regular terms as
      $\varepsilon\to 0$}\right) \nonumber \\
  ={}& \frac{S_b(\alpha_t)^2 S_b(\varepsilon)^2
    \mathrm{Res}\left(S_b^2,0\right)}{S_b(2\alpha +\alpha_t -
    Q+\varepsilon) S_b(Q+\alpha_t-2\alpha+\varepsilon)} + \dotsb\,
  . \label{eq:int-result}
\end{align}
Here we've used \eqref{eq:sbinv} to relate the coefficient of the zero
at $S_b(Q)^2$ to the residue $\mathrm{Res}\left(S_b^2,0\right)$, which
is given in equation~(\ref{eq:res-sb2}) (though we won't need its
explicit value). As the prefactor multiplying this integral vanishes,
we also won't need the explicit expression for the regular part of
this integral.

Combining the prefactor \eqref{eq:pref-sid} with the integral
\eqref{eq:int-result}, we obtain the following for the crossing kernel
(up to unimportant constant factors):
\begin{align}
  F_{0, \alpha_t}
  \begin{bmatrix} \displaystyle \alpha & \alpha \\ \alpha & \alpha
  \end{bmatrix}
  \propto{}& \mu(\alpha_t) \left( \frac{\Gamma_b(2\bar \alpha -
      \alpha_t) \Gamma_b(\alpha_t)}{\Gamma_b(2\bar\alpha)}\right)^2
  \frac{\Gamma_b(2Q) \Gamma_b(Q)}{ \Gamma_b(2\alpha_t)
    \Gamma_b(2\bar\alpha_t)} 
  \frac{S_b(2\alpha - \bar \alpha_t)}{S_b(2\alpha) S_b(\alpha_t)}
  \mathrm{Res}\left(S_b^2,0\right)\, .
\end{align}
For generic fixed $\alpha$, this result is a non-trivial meromorphic
function of $\alpha_t$, which indicates that there will be a continuous
contribution to the $t$-channel decomposition. However, when the
external operator is degenerate, either $\alpha = \alpha_{r,s}$ or
$\bar \alpha = \alpha_{r,s}$, $S_b(2\alpha)$ or $\Gamma_b(2 \bar
\alpha)$ diverges and therefore the crossing kernel vanishes unless
$\alpha_t$ takes particular values to cancel these zeros. This
demonstrates exactly what we set out to show. A necessary condition
for a finite operator product expansion for $[\phi] \times [\phi]$ is
that the $s$-channel identity block decomposes into a discrete sum in
the $t$-channel,\footnote{This is assuming that $[\phi] \times [\phi]
  \supset [1]$, which excludes the second family of solutions found in
  section 
    \ref{sec:one-plus-eps}. However, this peculiar case can
  be ruled out if we demand that $\phi$ have a non-zero two point
  function.} and here we've seen that this condition is enough to
imply that $\phi$ must be a degenerate operator.

\section*{Acknowledgments}

We would like to thank Ethan Dyer, Jared Kaplan, Leonardo Rastelli, Stephen Shenker, Herman Verlinde, and Xi Yin for valuable discussions.  We especially thank Stephen Shenker for suggesting to search for finite closed sub-algebras with the bootstrap, and to Ethan Dyer for collaboration during some early stages.  ALF is supported by the US Department of Energy Office of Science under Award Number DE-SC-0010025. We would also like to thank the GGI in Florence for hospitality as this work was completed.

\begin{appendices}

\section{Minimal Model Operators Near the Edge of the Kac Table}
\label{app:minmodtrunc}

In this appendix, we recall some relevant facts about the minimal models. In particular, we review which operators have an OPE algebra that truncates as (\ref{eq:ansatz1}). 

  The $\mathcal{M}(p,p')$ minimal model is defined by central charge $c$ and operators with weight $h=h_{r,s}$ given by \cite{DiFrancesco:1997nk} 					
	\begin{align}
	c&=1-\frac{6(p-p')^2}{pp'}, \label{eq:c_min_model} \\
	h_{r,s}&=\frac{(pr-p's)^2-(p-p')^2}{4pp'}, \quad 1\leq r < p', \quad 1\leq s< p. \label{eq:h_rs}
	\end{align}
Here $p$, $p'$ are coprime integers with $p>p'$. We denote primary operators with $h=h_{r,s}$ by $\Phi_{(r,s)}$. The Verma module $V(c,h_{r,s})$ generated by $\Phi_{(r,s)}$ then has the first null vector at level $rs$. The presence of an infinite cascade of null states forces the OPE in minimal model theories to truncate to a finite number of operators,  with the general fusion rules in the $\mathcal{M}(p,p')$ minimal model given by
	\begin{equation}
	\begin{aligned}[c]
	[\Phi_{(r,s)}]\times[\Phi_{(m,n)}]&=\sum_{\substack{k=1+|r-m|\\k+r+m=1\bmod 2}}^{k_{max}}\sum_{\substack{l=1+|s-n|\\l+s+n=1\bmod 2}}^{l_{max}}[\Phi_{(k,l)}], \\
	k_{max}&=\min(r+m-1,2p'-1-r-m), \\
	l_{max}&=\min(s+n-1,2p-1-s-n).
	\end{aligned}
	\end{equation}
Note the above sums are incremented by two. 

We are interested in the case $(r,s)=(m,n)$, for which the above becomes
	\begin{equation}
	\begin{aligned}[c]
	[\Phi_{(r,s)}]\times[\Phi_{(r,s)}]& = \sum_{\substack{k=1\\k+2r=1\bmod 2}}^{k_{max}}\sum_{\substack{l=1\\l+2s=1\bmod 2}}^{l_{max}}[\Phi_{(k,l)}],  \\
	k_{max}&=\min(2r-1,2p'-2r-1), \\
	l_{max}&=\min(2s-1,2p-2s-1).
	\end{aligned}
	\end{equation}
If we want the OPE to again contain operator $\Phi_{(r,s)}$ it must be the case that both $r$ and $s$ are odd integers. This is because the sums are incremented by odd integers. Then the OPE will have the form
	\begin{equation}
	[\Phi_{(r,s)}]\times[\Phi_{(r,s)}] = [1]+[\Phi_{(1,3)}]+[\Phi_{(3,1)}]+\dotsb+[\Phi_{(r,s)}]+\dotsb+[\Phi_{(k_{max},l_{max})}].
	\end{equation}
Demanding $(k_{max},l_{max})=(r,s)$ so that the algebra truncates to (\ref{eq:ansatz0}), (\ref{eq:ansatz1}), or (\ref{eq:ansatz2}), we recover the possibilities mentioned in the text (\ref{eq:opetrunc01}),(\ref{eq:opetrunc2}).

\section{Special functions}
\label{sec:funcs}
The function $\Gamma_b(x)$ is defined in terms of the Barnes double
Gamma function $\Gamma_2(x\vert \omega_1, \omega_2)$
\cite{barnes,Nakayama} as:
\begin{align}
  \Gamma_b(x) ={}& \frac{\Gamma_2(x \vert b,
    b^{-1})}{\Gamma_2(Q/2 \vert b, b^{-1})}\, , & Q ={}& b
  +b^{-1}\, .
\end{align}
Since $\Gamma_2(x \vert \omega_1, \omega_2)$ is symmetric under
exchange of $\omega_1$ and $\omega_2$, $\Gamma_b(x) =
\Gamma_{b^{-1}}(x)$. An important property of $\Gamma_b(x)$ is the
following shift relation:
\begin{align}
  \frac{\Gamma_b(x+b^{\pm 1})}{\Gamma_b(x)} ={}& \frac{\sqrt{2
      \pi}}{\Gamma(b^{\pm 1} x)} b^{\pm (x b^{\pm 1} - \frac{1}{2})}\, .
  \label{eq:gammabshift}
\end{align}
Using this relation and the values $\Gamma_2(b^{\pm 1} \vert b,
b^{-1}) = \sqrt{2\pi b^{\pm 1}}$, $\Gamma_b(nb+mb^{-1})$ can be
evaluated in terms of $\Gamma_2(Q/2 \vert b, b^{-1})$ for any positive
integers $n,m$. $\Gamma_b(x)$ is a non-vanishing meromorphic function
with poles at:
\begin{align}
  \Gamma_b(x)^{-1} = 0 \qquad \Leftrightarrow \qquad x ={-} nb
  -mb^{-1}\, , \quad n,m \in \mathbb{Z}^{\geq 0}\, ,
  \label{eq:gammabpoles}
\end{align}
and the residue at $x=0$ is:
\begin{align}
  \mathrm{Res}(\Gamma_b,0) = \frac{1}{\Gamma_2(Q/2 \vert b, b^{-1})}\,
  . \label{eq:gammabres0}
\end{align}
The shift relation \eqref{eq:gammabshift} then fixes the residue at
all other poles. Near $x=0$, $\Gamma_b$ can be expanded as:
\begin{align}
  \Gamma_2(Q/2 \vert b,b^{-1}) \Gamma_b(x) ={}& \frac{1}{x} -
  \gamma_{22}(b) + {\cal O}(x)\, , \label{eq:gammab-exp}
\end{align}
where $\gamma_{22}(b)$ is given by \cite{barnes,spreafico}:
\begin{align}
  \gamma_{22}(b) ={}& \frac{1}{b} \left(1 + \frac{b^2}{2} \right)
  \gamma - \frac{1}{2b} \log 2\pi + \frac{1}{2b} \left( 1 - b^2
  \right) \log b^{-1} - \frac{1}{b} \log b \nonumber \\
  &\quad - i b \int_0^\infty \mathrm{d} y \, \frac{\psi(1+i b^2 y) -
    \psi (1 - i b^2 y)}{e^{2\pi y} -1}\, . \label{eq:gamma22-def}
\end{align} 
Here $\gamma$ is the Euler-Mascheroni constant and $\psi$ is the
digamma function $\psi(x) = \frac{\Gamma'(x)}{\Gamma(x)}$.

The double sine function $S_b(x)$ is defined in terms of $\Gamma_b$
as:
\begin{align}
  S_b(x) ={}& \frac{\Gamma_b(x)}{\Gamma_b(Q-x)}\, . \label{eq:sbdef}
\end{align}
An immediate consequence of this definition is:
\begin{align}
  S_b(x) S_b(Q-x) ={}& 1\, . \label{eq:sbinv}
\end{align}
\eqref{eq:gammabpoles} implies that $S_b$ has the
following poles and zeros:
\begin{align}
  \text{ Poles: }{}& x = {-} (nb+m b^{-1})\, , & \text{Zeros: }& x = Q
  + nb + mb^{-1}\, , & n,m \in{}& \mathbb{Z}^{\geq 0}\,
  , \label{eq:sbpoles}
\end{align}
and \eqref{eq:gammabshift} gives the following shift relation:
\begin{align}
  \frac{S_b(x+b^{\pm 1})}{S_b(x)} ={}& 2 \sin(\pi b^{\pm 1} x)\,
  . \label{eq:sbshift}
\end{align}
This relation and the values of $\Gamma_b(b^{\pm 1})$ fix the residue
$\mathrm{Res}(S_b,0) = (2\pi)^{-1}$; further application of the shifts
\eqref{eq:sbshift} determine all other residues:
\begin{align}
  \mathrm{Res}\left(S_b,-nb-mb^{-1} \right) ={}& \frac{(-1)^{n+m + nm}}{2\pi}
  \left[ \prod_{k=1}^n 2 \sin(k \pi b^2) \prod_{l=1}^m 2 \sin(l \pi
    b^{-2}) \right]^{-1}\, .
  \label{eq:sbres}
\end{align}
The inversion property \eqref{eq:sbinv} then implies that the
coefficient of the zero at $x=Q+nb+mb^{-1}$ is
$[\mathrm{Res}(S_b,Q-x)]^{-1}$. If we define the q-numbers
\begin{align}
  [n] ={}& \frac{\sin(\pi b^2 n)}{\sin(\pi b^2)}\, , & [m]' ={}&
  \frac{\sin(\pi b^{-2} m)}{\sin (\pi b^{-2})}\, , \label{eq:qnum-def}
\end{align}
and the corresponding q-factorials $[n]! =
[n][n-1]!$, $[m]'!=[m]'[m-1]'!$, we can rewrite \eqref{eq:sbres} as:
\begin{align}
  \mathrm{Res}\left(S_b,-nb-mb^{-1} \right) ={}& \frac{(-1)^{n+m + nm}}{2\pi
    [n]! [m]'!}  \left(2 \sin \pi b^2 \right)^n \left(2 \sin \pi
    b^{-2} \right)^m\, . \label{eq:sbresqfac}
\end{align}

We will also need the residue of $S_b^2(x)$ at $x=0$, which is a
double pole. To evaluate this residue, we use \eqref{eq:sbshift} to
rewrite $S_b(x)$ as:
\begin{align}
  S_b(x) ={}& \frac{\Gamma(1-bx)\Gamma({-}x/b)}{2\pi b^{1-x(b-1/b)}}
  \frac{\Gamma_b(x)}{\Gamma_b({-}x)}\, . \label{eq:sb-rewrite}
\end{align}
Using the series expansion \eqref{eq:gammab-exp}, we find:
\begin{align}
  S_b(x)^2 ={}& \frac{1}{4\pi^2 x^2} + \frac{(1+b^2)\gamma - 2 b
    \gamma_{22} - (1-b^2) \log b}{2 \pi^2 b x} + {\cal O}(x^0)\, ,
\end{align}
so the residue is:
\begin{align}
  \mathrm{Res}(S_b^2,0) ={}& \frac{(1+b^2)\gamma - 2 b \gamma_{22} - (1-b^2)
    \log b}{2 \pi^2 b}\, . \label{eq:res-sb2}
\end{align}

\section{Coulomb gas and minimal model fusion matrices}
\label{sec:coulomb-gas}

The Coulomb gas is a standard technique to compute correlation functions of
degenerate operators with general central charge $c$ using a modified
free scalar CFT, where the degenerate operators are realized as
exponentials of the free scalar \cite{DiFrancesco:1997nk}.    While we emphasize that the bootstrap methods in this paper do not rely on imposing any degeneracy condition, we do need to be able compare to minimal models in order to see whether all of our solutions turn out to be minimal models or not.  Furthermore, the Coulomb gas formalism is directly connected to the methods of \cite{Teschner1,Teschner2,Teschner3} for the crossing matrices.  In this appendix, we will review a few very basic elements of this formalism.

The basic idea is to 
consider the theory of a scalar with the standard OPE
\begin{equation}
  \label{eq:app-CG-phi-OPE}
  \phi(z, \bar z) \phi(w,\bar w) \sim {-} \alpha' \log \vert z-w \vert\, ,
\end{equation}
but with a modified stress tensor
\begin{equation}
  \label{eq:app-CG-T-def}
  T(z) = {-} \frac{1}{\alpha'} (\partial \phi)^2(z) + Q \partial^2  \phi(z)\, ,
\end{equation}
 obtained from the action
\begin{equation}
  \label{eq:app-CG-action}
  S = \frac{1}{4\pi \alpha'} \int d^2 z \, g^{1/2} \left\{
    g^{ab} \partial_a \phi \partial_b \phi + \alpha' Q \phi R
  \right\}\, .
\end{equation}
The Ricci term in the action
 contributes to, among
other things, global symmetries and Ward identities as well as the
central charge:
\begin{align}
  \label{eq:app-CG-central-charge}
  c = 1 + 6 \alpha' Q^2\, .
\end{align}

Conformal primaries of interest are the
vertex operators
\begin{align}
  V_\alpha(z,\bar z) \sim e^{i \alpha \phi(z,\bar z)}\, ,
\end{align}
suitably regularized.
Since the propagator
for $\phi(z,\bar z)$ is that of a free boson, it is straightforward to
calculate arbitrary correlation functions of the vertex operators.
\begin{equation}
  \label{eq:app-CG-Vs-cf}
  \left\langle \prod_{i=1}^n V_{\alpha_i}(z_i) \right\rangle_{Q}
  \simeq \prod_{i<j} z_{ij}^{\frac{\alpha'}{2} \alpha_i \alpha_j}\, .
\end{equation}

The subscript $Q$ denotes that these correlation functions are
actually only non-zero if a neutrality condition is satisfied. The
Ricci scalar term in the action modifies the nature of the global
symmetry $\phi \to \phi+a$, effectively placing a background charge of
${-} 2i Q$ at infinity. More precisely, despite the fact that the
Ricci scalar coupling `breaks' the shift symmetry, a modified Ward
identity survives that forces non-zero correlation functions to have
total charge ${-} 2i Q$.\footnote{This follows from the fact that
  $\int d^2 z \, g^{1/2} R$ measures the Euler number and hence is a
  topological invariant.} Taking $Q = i \alpha_0$ and noting that the
vertex operators $V_\alpha(z)$ have charge $\alpha$ under this global
symmetry, the neutrality conditions reads:
\begin{align}
  \sum_i \alpha_i ={}& 2 \alpha_0\, .
  \label{eq:neut}
\end{align}
In the case of a two point function, this prescription yields
\begin{align}
  \langle V_\alpha(z) V_{2\alpha_0 - \alpha}(0) \rangle \sim
  z^{\frac{\alpha'}{2} \alpha (2\alpha_0 - \alpha)} \, ,
\end{align}
which implies that one should take $V_\alpha^\dagger = V_{2\alpha_0 - \alpha}$, and 
\begin{align}
  h_\alpha ={}& \frac{\alpha'}{4} \alpha (\alpha - 2 \alpha_0)\,
  .
\end{align}

In order to generalize the set of correlators that can be non-vanishing consistently with the neutrality condition (\ref{eq:neut}), one adds in non-local
`screening charges', which are conformally
invariant operators that soak up extra charge:
\begin{align}
  Q_\alpha ={}& \oint_C d z V_\alpha(z)\, .
\end{align}
For this to be conformally invariant, the vertex operator must have
weight $1$ to offset the measure, which requires:
\begin{align}
  \alpha(\alpha - 2 \alpha_0) ={}& \frac{4}{\alpha'} \qquad
  \Longrightarrow \qquad \alpha_\pm =\alpha_0 \pm \sqrt{\alpha_0^2 +
    4/\alpha'}\, .
\end{align}
Inserting such an operator does not affect the conformal
Ward identities.\footnote{To see this, one uses the fact that
\begin{equation}
  [L_n, V_\alpha(z)] = \left[z^{n+1} \partial_z + \Delta_\alpha (n+1) z^n \right] V_\alpha\, .
\end{equation}
If $\Delta_\alpha = 1$, i.e.~$\alpha = \alpha_\pm$, then this is equivalent to
\begin{equation}
  [L_n, V_\alpha(z)] = \left[z^{n+1} \partial_z + (n+1) z^n \right] V_\alpha(z) = \partial_z \left[ z^{n+1} V_\alpha(z) \right]\, .
\end{equation}
Provided the operator $V_\alpha(z)$ takes the same value at the beginning and end of the integration contour, integration of the above equation implies $[L_n, Q] = 0$.  As we will see a little later on, the integration contours are chosen to satisfy this constraint.
}  Therefore, this is a constructive method for generating correlation functions that are consistent with crossing symmetry and conformal symmetry, which for minimal models uniquely determines the correlation functions.

For simplicity and to make contact with more standard CFT notation, we
will set:
\begin{align}
  \alpha' = 4\, . 
\end{align}
Thus, the screening charges take the form:
\begin{align}
  Q_\pm ={}& \oint_C d z \, V_{\alpha_\pm}(z)\, , & \alpha_\pm ={}&
  \alpha_0 \pm \sqrt{\alpha_0^2 + 1}\, .
\end{align}

Some useful things to note about $\alpha_\pm$ are:
\begin{align}
  \alpha_+ + \alpha_- ={}& 2 \alpha_0\, , & \alpha_+ \alpha_- ={}& {-}
  1\, .
\end{align}
For later use, we define the parameters
\begin{align}
  \rho ={}& \alpha_+^2 \, , & \rho' = \alpha_-^2 = \frac{1}{\rho}\, .
\end{align}

To evaluate, say, the four-point function $\langle V_\alpha V_\alpha
V_\alpha V_{2\alpha_0 - \alpha} \rangle$, one must be able to add in factors of $Q_\pm$ to bring the total charge to $2 \alpha_0$.  If $2\alpha$ is a linear combination of
$\alpha_\pm$, i.e. if
\begin{align}
  2 \alpha ={}& (1- r) \alpha_+ + (1- s) \alpha_-\, ,
\end{align}
then one can consider
\begin{align}
  \langle V_\alpha V_\alpha V_\alpha V_{2\alpha_0 - \alpha} Q_+^{r-1}
  Q_-^{s-1} \rangle\, .
\end{align}
By construction, the operators in this correlation function satisfy
the neutrality condition.

It is conventional to  parametrize these nice charge values of $\alpha$ by
\begin{align}
  \alpha_{r,s} \equiv{}& \frac{1-r}{2} \alpha_+ + \frac{1-s}{2}
  \alpha_- = \alpha_0 - \frac{1}{2} (r \alpha_+ + s \alpha_-)\, ,
\end{align}
corresponding to dimensions of
\begin{align}
  h_{r,s} ={}& \alpha_{r,s} (\alpha_{r,s} - 2 \alpha_0) = {-}
  \alpha_{r,s} \alpha_{{-} r, {-} s} = \frac{(r \alpha_+ + s
    \alpha_-)^2}{4} - \alpha_0^2\, ,
\end{align}
which are the usual degenerate conformal weights. 

Similar considerations apply to  correlation functions with more
than one operator, i.e. $
  \langle V_{\alpha_1} V_{\alpha_2} V_{\alpha_3} V_{2\alpha_0 -
    \alpha_4} Q_+^r Q_-^s \rangle$.

The above Coulomb gas formalism produces integral representations of the correlators in minimal models. For instance, the correlator 
$
  F(z_i) \equiv \langle V_{1,2}(z_1) V_{1,2}(z_2) V_{r,s}(z_3) V_{{-} r,{-} s}(z_4)
  Q_- \rangle\, $
can be represented as
\begin{align}
  F(z_i) ={}& \oint_C d u \, \langle V_{1,2}(z_1) V_{1,2}(z_2)
  V_{r,s}(z_3) V_{{-} r,{-} s}(z_4) V_-(u) \rangle \nonumber \\
  ={}& z_{12}^{2 \alpha_{1,2}^2} (z_{13} z_{23})^{2 \alpha_{1,2}
    \alpha_{r,s}} (z_{14} z_{24})^{2\alpha_{1,2} \alpha_{{-} r, {-}
      s}} z_{34}^{{-} 2\Delta_{r,s}} \times \nonumber \\
  &\quad \oint_C d u \, [(z_1 - u)(z_2 - u)]^{2 \alpha_{1,2}
    \alpha_-} (z_3 - u)^{2 \alpha_{r,s} \alpha_-} (z_4 - u)^{2
    \alpha_{{-} r, {-} s} \alpha_-}\, .
\end{align}
Using global conformal invariance to send
$z_1 \to \infty$, $z_2 \to 1$, $z_3 \to z$ and $z_4 \to 0$, this reduces to 
\begin{align}
  F(z) ={}& (1-z)^{2\alpha_{1,2} \alpha_{r,s}} z^{2 \alpha_{1,2}^2}
  \oint d u\, u^{2 \alpha_{1,2} \alpha_-} (u-z)^{2\alpha_{1,2}
    \alpha_-} (u-1)^{2\alpha_{r,s} \alpha_-} \, ,
\end{align}
up to some phase factors that we will fix independently.
This integral depends on the choice of contour. This contour should be
single valued, that is the integrand should be single valued upon
going around the entire contour, while also enclosing at least one
singular point so that it is non-vanishing. A slick way to do so is to
use the Pochhammer contour, which encloses two of the singularities
twice, once clockwise and once counter clockwise. Since any monodromy
obtained by going around a singularity is eventually cancelled by
going around in the opposite direction, the integrand is single
valued. Furthermore, by collapsing the contour to the line connecting
the singularities, the integral reduces to a single integral between
the two singular points, though there is a phase that one has keep
track of. In any case, there are two independent such contours, which
correspond to the two different conformal blocks that are allowed in
the OPE of $V_{1,2} \times V_{1,2}$.  In the present case, they have
simple representations as hypergeometric functions, via the identities
\begin{align}
  \int_1^\infty d u\, u^a (u-1)^b (u-z)^c ={}& I_1(a,b,c;z)
  \\
  ={}& \frac{\Gamma({-} a - b - c - 1) \Gamma(b+1)}{\Gamma({-} a - c)}
  {}_2F_1({-} c, {-} a- b - c - 1; {-} a - c;z)\, , \nonumber \\
  \int_0^z d u\, u^a (1-u)^b (z-u)^c ={}& I_2(a,b,c;z)
  = z^{1+a+c} \int_0^1 d u\, u^a (1-u)^c (1- z u)^b \nonumber \\
  ={}& z^{1+a+c} \frac{\Gamma(a+ 1) \Gamma(c+1)}{\Gamma(a+c+2)}
  {}_2F_1(a+1,{-} b; a+c+2;z)\, .
\end{align}

The generalization to higher level degenerate operators is
straightforward, if tedious. For a set of four external operators with
charges $\alpha_i =\alpha_{r_i,s_i}$, for $i=1,2,3$ and $\bar \alpha_4
=2\alpha_0 - \alpha_{r_4,s_4}$, one adds in a factor of $Q_+^{m-1}
Q_-^{n-1}$, where
\begin{align}
  m ={}& \frac{r_1+r_2+r_3-r_4}{2} \, , & n ={}& \frac{s_1 + s_2 +
    s_3 - s_4}{2}\, .
\end{align}
This leads to an $(m-1)(n-1)$-fold integral expression for the
(holomorphic) correlation function. For each integral there are two
independent contour choices and this leads to a total of $M=mn$
independent analytic functions ${\cal F}_i$ where $i=1,\dotsc, M$. For
generic $m$ and $n$, these analytic functions cannot be explicitly
given in terms of special functions as for the $m=1$, $n=2$ case
above, but the monodromy properties are readily obtained via contour
manipulation.

So far we have concentrated on the holomorphic correlation functions,
but for a physical theory we must construct add in the
anti-holomorphic sector. Restricting to the case of scalar primaries,
we can construct the physical correlation function as
\begin{equation}
  G(z,\bar z) = \sum_{k,l=1}^M C_{kl} {\cal F}_k(z) \overline{{\cal F}_l(z)}\, .
\end{equation}
To specify the matrix $C_{kl}$, we require that the physical
correlation function be single valued and hence monodromy free. In
particular we check the monodromy around $z=0$ and $z=1$. The $z=0$
case is simple and forces $C_{kl}$ to be diagonal $C_{kl} = C_k
\delta_{kl}$:
\begin{equation}
  G(z, \bar z) = \sum_k C_k \left\vert {\cal F}_k(z) \right\vert^2\, .
\end{equation}

The $z=1$ monodromy is much more involved. The approach worked out  in
\cite{Dotsenko:1984nm,Dotsenko:1984ad} is to use the integral expressions to rewrite the ${\cal
  F}_k(z)$ in terms of $M$ new analytic functions $\tilde {\cal
  F}_k(z)$ with diagonal monodromy around $z=1$; physically this
procedure is expressing the conformal blocks in the $s$-channel in
terms of the $t$-channel blocks:
\begin{equation}
  {\cal F}_k(z) = F\begin{bmatrix} \displaystyle
    \alpha_2 & \alpha_3\\ \alpha_1 & \alpha_4
  \end{bmatrix}_{kl} \tilde {\cal F}_l(z)\, .
\end{equation}
In terms of the $t$-channel blocks, the correlation function reads
\begin{equation}
  G(z, \bar z) = \sum_{k,l,m} C_k F\begin{bmatrix} \displaystyle
    \alpha_2 & \alpha_3 \\ \alpha_1 & \alpha_4
  \end{bmatrix}_{kl} F\begin{bmatrix} \displaystyle
    \alpha_2 & \alpha_3 \\ \alpha_1 & \alpha_4
  \end{bmatrix}_{km}^* \tilde {\cal F}_l(z) \overline{\tilde {\cal
      F}_m(z)} \equiv \sum_{l,m} \tilde C_{lm} \tilde {\cal F}_l(z)
  \overline{\tilde {\cal F}_m(z)}\, .
\end{equation}
Therefore diagonal monodromy around $z=1$ requires $\tilde C_{lm} = 0$
for $l \neq m$. With this constraint, one can solve for the
coefficients $C_k$ up to an overall coefficient\footnote{This solution
  follows from multiplying $\tilde C_{lm} = 0$ by $(F^{-1})_{ln}$ and
  summing for all $l\neq m$. This yields $C_n F^*_{nm} = \tilde C_{mm}
  (F^{-1})_{mn}$, and taking the ratio of the $n=k$, $m=M$ and $n=m=M$
  equations produces the given solution. Note that though it seems
  like we have a substantially overconstrained system of equations
  $\tilde C_{lm}=0$ for the $M$ unknowns $C_k$, its solvability is
  guaranteed as it arises as a monodromy matrix of a linear
  differential equation.}
\begin{equation}
  \frac{C_k}{C_M} = \frac{F_{MM}^* (F^{{-}1})_{Mk}}{F_{kM}^* (F^{{-}1})_{MM}}\, .
\end{equation}
Provided we normalize the blocks ${\cal F}_k(z)$ appropriately, the
$C_k$ are nothing but the OPE coefficients. We give the explicit
solutions for these OPE coefficients in the next appendix.

Finally, we give the closed form expressions for the fusion matrix
$F$\footnote{While the particular matrix elements needed to determine
  the OPE coefficients were evaluated in
  \cite{Dotsenko:1984nm,Dotsenko:1984ad}, it seems the general fusion
  matrix was not obtained until later \cite{furlan,Hou}.}
\begin{equation}
  \label{eq:deg-fus-matrix}
  F
  \begin{bmatrix}
    \alpha_2 & \alpha_3 \\
    \alpha_1 & \alpha_4
  \end{bmatrix}_{(p_s,p_s'),(q_t,q_t')} = \frac{
    N^{(m,n)}_{k_2,k_2'}(b,a,c,d;\rho)}{N^{(m,n)}_{k_1,k_1'}(a,b,c,d;\rho)}
  \alpha^{(m)}_{k_1,k_2}(a,b,c,d;\rho)
  \alpha^{(n)}_{k_1',k_2'}(a',b',c',d';\rho') \, ,
\end{equation}
where the parameters are defined as
\begin{align}
  a={}& 2 \alpha_+ \alpha_1, & b={}& 2 \alpha_+ \alpha_3, & c={}& 2
  \alpha_+ \alpha_2\, , & d ={}& 2 \alpha_+ \bar \alpha_4 \,
  , & \rho ={}& \alpha_+^2\, , \\
  a'={}& 2 \alpha_- \alpha_1 , & b'={}& 2 \alpha_- \alpha_3 , & c'={}&
  2 \alpha_- \alpha_2 \, , & d' ={}& 2 \alpha_- \bar \alpha_4 \, , &
  \rho' ={}& \alpha_-^2 = 1/\rho\, ,
\end{align}
the indices as ($k_i, k_i'$, $i=1,2$, are simply convenient
parametrizations for the exchanged operators and recall $m,n$ are the
number of screening charges required)
\begin{align}
  m ={}& \frac{r_1+r_2+r_3-r_4}{2}\, , & n ={}&
  \frac{s_1+s_2+s_3-s_4}{2}\, , \\
  k_1 ={}& \frac{r_1 + r_2 + 1 - p_s}{2}\,, & k_1' ={}& \frac{s_1 + s_2 +
    1 - p_s'}{2}\,, \\
  k_2 ={}& \frac{r_2 + r_3 + 1 - p_t}{2}\,, & k_2' ={}& \frac{s_2 + s_3 +
    1 - p_t'}{2}\,,
\end{align}
the normalization functions as (note that $a' = {-}a/\rho$,
$b'={-}b/\rho$, etc.~)
\begin{align}
  N^{(m,n)}_{p,p'}(a,b,c,d;\rho) ={}& {\cal J}_{m-p,n-p'}(d,b;\rho)
  {\cal J}_{p-1,p'-1}(a,c;\rho)\, , \\
  {\cal J}_{p,q}(a,b;\rho) ={}& \rho^{2 pq} \prod_{i,j=1}^{p,q}
  \frac{1}{i \rho - j}\prod_{i=1}^p \frac{\Gamma(i
    \rho)}{\Gamma(\rho)} \prod_{j=1}^q \frac{\Gamma(j
    \rho')}{\Gamma(\rho')} \\
  &\times \prod_{i=0}^{p-1} \frac{\Gamma(1+a+i \rho) \Gamma(1+b+i
    \rho)}{\Gamma(2-2q+a+b+(p-1+i)\rho)}
  \nonumber\\
  &\times \prod_{j=0}^{q-1} \frac{\Gamma(1+a'+j \rho') \Gamma(1+b'+j
    \rho')}{\Gamma(2-2p+a'+b'+(q-1+j)\rho')}
  \nonumber \\
  &\times \prod_{i,j=0}^{p-1,q-1} \frac{1}{(a+i \rho - j) (b + i \rho
    - j)[a + b + \rho(p-1+i) - (q-1+j)]}\, , \nonumber
\end{align}
and finally
\begin{align}
  \alpha^{(m)}_{j,k}(a,b,c,d;\rho) ={}&
  \sum_{p=\max(j,k)}^{\min(m,j+k-1)} \prod_{i=1}^{p-j}
  \frac{s[(j+k+i-p-1)\rho]}{s(i\rho)} \prod_{i=1}^{m-p}
  \frac{s[(p-k+i)\rho]}{s(i \rho)} \\
  &\quad \frac{\prod_{i=0}^{m-p-1} s[1+a+(j-1+i) \rho]
    \prod_{i=0}^{p-k-1}s[1+d+(m-j+i)\rho]}{\prod_{i=0}^{m-k-1}
    s[a+d+(m-k-1+i)\rho]} \nonumber \\
  &\quad \frac{\prod_{i=0}^{j+k-p-2} s[1+b+(m-j+i) \rho]
    \prod_{i=0}^{p-j-1}s[1+c+(j-1+i)\rho]}{\prod_{i=0}^{k-2}
    s[b+c+(k-2+i)\rho]} \, ,  \nonumber \\
  s(x) ={}& \sin (\pi x)\, .
\end{align}

As an aside, we note that the fusion matrix can be interprated as the
product of Racah-Wigner symbols (closely related to the $6$J symbols)
for the quantum group $U_q(su(2))$. This correspondence can be
motivated by recalling the coset construction of the minimal models,
e.g.~the $m$th CFT in the unitary series can be defined as the coset
$SU(2)_{m+2} \times SU(2)_1/ SU(2)_{m+3}$. The factorization of the
fusion matrix (into $\alpha^{(m)}_{k_1,k_2}$ and
$\alpha^{(n)}_{k_1',k_2'}$) is then due to the factors $SU(2)_{m+2}$
and $SU(2)_{m+3}$ in the coset (the $SU(2)_1$ factor, being at level
$1$, behaves trivially in the field identification between the coset
and the minimal models). The precise correspondence (restricting for
the moment to the unitary series) can be written as
\begin{align}
  \alpha^{(m)}_{j,k}(a,b,c;\rho) ={}&
  ({-}1)^{(j-1)(1+r_{23})+(k-1)(1+r_{12})+(m-1)r_{123} + j_{12} -
    j_{34}-2j_5} \nonumber \\
  & \times \frac{\sqrt{s((2j_5+1)\rho) s((2j_6+1)\rho)}}{s(\rho)}
  \sqrt{\frac{X^{(m)}_k(b,a,c;\rho)}{ X^{(m)}_j(a,b,c;\rho)}}
  \nonumber \\
  & \times
  \begin{pmatrix}
    j_1(a) & j_2(c) & j_5(a,c,j) \\ j_3(b) & j_4(d) & j_6(b,c,k)
  \end{pmatrix}_q^{RW}\,,
\end{align}
where
\begin{align}
  2j_i + 1 ={}& r_i\, , & 2j_5+1 ={}& p_s \,, & 2j_6 + 1 ={}& p_t \,
  ,& q ={}& e^{{-}i \pi \rho}\, ,
\end{align}
and a label with multiple subscripts denotes a summation over the
corresponding values, e.g.~$r_{ij} = r_i + r_j$, $r_{ijk} = r_i + r_j
+ r_k$, etc. The $X^{(m)}_j$ are normalization factors that can be
found in \cite{Dotsenko:1984ad} eq. (3.19), and the Racah-Wigner
symbols themselves can be found in \cite{kirillov}. We note that, with
some care regarding the normalization factors, the fusion matrices
\eqref{eq:deg-fus-matrix} can be obtained from the Liouville fusion
kernel \eqref{eq:tesch-fusion}, as shown in detail in
\cite{paulina-et-al}.

For clarity and comparison, below we explicitly show some of the fusion matrices that obtain from the above expression.  First, consider the simplest non-trivial case, a degenerate $\phi_{1,2}$ operator.  This closes to $\phi_{1,1} = \mathbf{1}$ and $\phi_{1,3}=\epsilon$.  In the Ising model, $m=3$, the resulting $2 \times 2$ crossing matrix is the only non-trivial one, and is just
\be
F &=& \left( \begin{array}{cc} \frac{1}{\sqrt{2}} & \frac{1}{2 \sqrt{2}} \\ \sqrt{2} & - \frac{1}{\sqrt{2}} \end{array} \right).
\ee
Actually, there is no reason that $m$ needs to be restricted to an integer.  We can take $m = -\frac{b^2}{1+ b^2}$ with $b$ arbitrary, and obtain
\be
F &=& \left(
\begin{array}{cc}
 -\frac{1}{2} \sec \left(b^2 \pi \right) & -\frac{\csc \left(2 b^2 \pi \right) \Gamma
   \left(-3 b^2-1\right) \Gamma \left(2 b^2+2\right) \sin \left(3 b^2 \pi
   \right)}{\Gamma \left(-2 b^2\right) \Gamma \left(b^2+1\right)} \\
 -\frac{\Gamma \left(-2 b^2\right) \Gamma \left(b^2+1\right) \sec \left(b^2 \pi
   \right)}{2 \Gamma \left(-3 b^2-1\right) \Gamma \left(2 b^2+2\right)} & \frac{1}{2}
   \sec \left(b^2 \pi \right). \\
\end{array}
\right)
\ee
For a more complicated example, consider the $m=4$ ($c=\frac{7}{10}$, tricritical Ising model) minimal model.  The most complicated OPE is that of the $\phi_{2,2} = \sigma$ operator, which contains $\phi_{1,1} = \mathbf{1}, \phi_{1,3} = \epsilon', \phi_{3,1} = \epsilon''$, and $\phi_{3,3} = \epsilon$.  The $\sigma$ four-point function therefore has a $ 4 \times 4 $ crossing matrix, given by
\be
F &=&
\left(
\begin{array}{cccc}
 \frac{\sqrt{3-\sqrt{5}}}{2} & \frac{\Gamma \left(\frac{2}{5}\right)^2}{50 \sqrt{2} \Gamma
   \left(\frac{6}{5}\right) \Gamma \left(\frac{8}{5}\right)} & \frac{\sqrt{3-\sqrt{5}}}{112} & -\frac{\Gamma
   \left(\frac{2}{5}\right)^2}{2 \sqrt{2} \Gamma \left(-\frac{2}{5}\right) \Gamma \left(\frac{6}{5}\right)} \\
 \frac{2 \sqrt{2} \left(-1+\sqrt{5}\right) \Gamma \left(\frac{6}{5}\right) \Gamma
   \left(\frac{8}{5}\right)}{\Gamma \left(\frac{7}{5}\right)^2} & -\frac{1}{2} \sqrt{3-\sqrt{5}} &
   \frac{\sqrt{\left(3-\sqrt{5}\right) \pi } \Gamma \left(\frac{8}{5}\right)}{8\ 2^{2/5} \Gamma
   \left(\frac{2}{5}\right) \Gamma \left(\frac{17}{10}\right)} & -3 \sqrt{3-\sqrt{5}} \\
 14 \sqrt{2} \left(-1+\sqrt{5}\right) & \frac{2\ 2^{9/10} \Gamma \left(\frac{2}{5}\right) \Gamma
   \left(\frac{17}{10}\right)}{\sqrt{\pi } \Gamma \left(\frac{8}{5}\right)} & -\frac{1}{2} \sqrt{3-\sqrt{5}} &
   \frac{25 \sqrt{2} \Gamma \left(\frac{2}{5}\right) \Gamma \left(\frac{12}{5}\right)}{\Gamma
   \left(-\frac{2}{5}\right) \Gamma \left(\frac{6}{5}\right)} \\
 -\frac{\left(-1+\sqrt{5}\right) \Gamma \left(-\frac{2}{5}\right) \Gamma \left(\frac{6}{5}\right)}{\sqrt{2}
   \Gamma \left(\frac{2}{5}\right)^2} & -\frac{1}{12} \sqrt{3-\sqrt{5}} & \frac{\sqrt{3-\sqrt{5}} \Gamma
   \left(-\frac{2}{5}\right) \Gamma \left(\frac{6}{5}\right)}{100 \Gamma \left(\frac{2}{5}\right) \Gamma
   \left(\frac{12}{5}\right)} & \frac{\sqrt{3-\sqrt{5}}}{2} \\
\end{array}
\right)\nn\\
\ee

\section{Minimal model OPE coefficients}
For completeness, here we present the OPE coefficients for the
Virasoro minimal models in closed form, first given in
\cite{Dotsenko:1984nm,Dotsenko:1984ad,Dotsenko:1985hi}.  These results
are obtained by analyzing the monodromies of the Coulomb gas integral
expressions for the conformal blocks discussed in the previous
appendix. For simplicity, we give only the coefficients for the
diagonal minimal models; for the calculation of the coefficients in
more general non-diagonal theories, see \cite{Fuchs,FuchsKlemm}.

In \cite{Dotsenko:1984nm,Dotsenko:1984ad,Dotsenko:1985hi}, it is shown
that the square of the OPE coefficients can be written as:
\begin{equation}
  \label{eq:app-mm-ope-eq}
  \left[C^{(r_1,s_1)}_{(r_2,s_2),(r_3,s_3)}\right]^2 = \frac{a(r_2,s_2)
    a(r_3,s_3)}{a(r_1,s_1)} \left[D^{(r_1,s_1)}_{(r_2,s_2),(r_3,s_3)}
  \right]^2\, ,
\end{equation}
where $a(r,s)$ and $D^{(r_1,s_1)}_{(r_2,s_2),(r_3,s_3)}$ are defined
as
\begin{align}
  a(r,s) ={}& \left[\prod_{i,j=1}^{s-1,r-1} \frac{1+i-\rho (1+j)}{i-j
      \rho}\right]^2 \left[\prod_{i=1}^{s-1} \frac{\Gamma(i \rho')
      \Gamma(2-\rho' (1+i))}{\Gamma(1-i \rho')\Gamma(\rho'(1+i)-1)}
  \right] \nonumber \\
  & \qquad \times \left[\prod_{j=1}^{r-1} \frac{\Gamma(j \rho)
      \Gamma(2-\rho
      (1+j))}{\Gamma(1-j \rho)\Gamma(\rho(1+j)-1)} \right]\, , \\
  D^{(r_1,s_1)}_{(r_2,s_2),(r_3,s_3)} ={}& \mu(l,l')
  \left[\prod_{i,j=0}^{l'-2,l-2} \tilde \lambda_{ij}(r_1,s_1)
      \lambda_{ij}(r_2,s_2)\lambda_{ij}(r_3,s_3) \right]
  \nonumber \\
  &\quad \qquad \times \left[\prod_{j=0}^{l-2} \tilde \tau_j(r_1,s_1;\rho)
  \tau_j(r_2,s_2;\rho) \tau_j(r_3,s_3;\rho)\right] \nonumber \\
  &\quad \qquad \times \left[\prod_{i=0}^{l'-2} \tilde \tau_i(s_1,r_1;\rho')
  \tau_i(s_2,r_2;\rho') \tau_i(s_3,r_3;\rho')\right]\, .  
\end{align}
Here $l=\frac{r_2+r_3-r_1+1}{2}$ and $l'=\frac{s_2+s_3-s_1+1}{2}$,
while the auxilliary functions $\mu$, $\lambda$, and $\tau$ are
defined as
\begin{align}
  \mu(l,l') ={}& \rho^{4(l-1)(l'-1)} \prod_{i,j=1}^{l'-1,l-1} (i-\rho
  j)^{{-}2} \prod_{i=1}^{l'-1}\frac{\Gamma(i \rho')}{\Gamma(1-i
    \rho')} \prod_{j=1}^{l-1}\frac{\Gamma(j \rho)}{\Gamma(1-j
    \rho)}\, , \\
  \lambda_{ij}(r,s) ={}& [(s - 1 - i) - \rho (r-1-j)]^{{-}2}\,, \\
  \tilde \lambda_{ij}(r,s) ={}& [(s + 1 + i) - \rho (r+1+j)]^{{-}2}\,, \\
  \tau_i(r,s;\rho) ={}& \frac{\Gamma(s-\rho
    (r-1-i))}{\Gamma(1-s+\rho(r-1-i))}\, , \\
  \tilde \tau_i(r,s;\rho) ={}& \frac{\Gamma(\rho
    (r+1+i)-s)}{\Gamma(1+s-\rho(r+1+i))}\, .
\end{align}
Finally, we recall that $\rho = 1/\rho' = \alpha_+^2$, which takes the
value $\rho= \frac{p}{q}$ for the minimal model ${\cal M}(p,q)$ (in
the notation of \cite{DiFrancesco:1997nk}). The unitary series corresponds
to $p=m+1$, $q=m$.

The correct use of the above expressions for the minimal models
requires a particular choice of indices $(r,s)$. In particular, if we
let $1 \leq r_2, r_3 \leq q-1$ and $1 \leq s_2,s_3 \leq p-1$, then we
must take
\begin{equation}
  \begin{split}
    r_1 \in{}& \left\{\vert r_2 - r_3 \vert + 1, \vert r_2 - r_3 \vert
      + 3, \dotsc, \min(r_2 + r_3 - 1, q-1 \text{ or } q -2)
    \right\}\, ,  \\
    s_2 \in{}& \left\{\vert s_2 - s_3 \vert + 1, \vert s_2 - s_3 \vert
      + 3, \dotsc, \min(s_2 + s_3 - 1, p-1 \text{ or } p-2) \right\}\,
    .
  \end{split}  
\end{equation}
For the last argument of the $\min$s, one is to take the terms with
the same remainder modulo $2$ as $\vert r_2 - r_3 \vert + 1$ and
$\vert s_2 - s_3 \vert + 1$. For example, if one wants to compute
$C^\epsilon_{\sigma\sigma}$ in the Ising model with $r_2 = r_3 = 1$
and $s_2 = s_3 = 2$, then the correct choice for $\epsilon$ is
$(r_1,s_1)=(1,3)$, as opposed to $(r_1, s_1) = (2,1)$. Furthermore,
the expression (\ref{eq:app-mm-ope-eq}) may give a non-zero answer
even if $(r_1,s_1)$ does not lie in the appropriate set; for example,
with $r_2 = r_3 = 1$ and $s_2 = s_3 = 2$, one finds
$C^{(1,2)}_{(1,2),(1,2)} \neq 0$ for generic $\rho$. This would imply
that $\sigma \subset \sigma \times \sigma$, which is clearly
false. Thus one can only use these results confidently once the
structure of the fusion algebra is known.

\end{appendices}

\newpage

\bibliographystyle{utphys}
\bibliography{OPEBib}

\begin{thebibliography}{10}
\ifx\href\asklfhas\newcommand{\href}[2]{#2}\fi
\ifx\arxivref\asklfhas\newcommand{\arxivref}[2]{\href{http://arxiv.org/abs/#1}{#2}}\fi
\ifx\doiref\asklfhas\newcommand{\doiref}[2]{\href{http://dx.doi.org/#1}{#2}}\fi
\parskip 0pt
\normalsize

\bibitem{Moore:1988uz}
G.~W. Moore \& N.~Seiberg,
\textit{``{Polynomial Equations for Rational Conformal Field Theories}''},
\doiref{10.1016/0370-2693(88)91796-0}{Phys.~Lett. \textbf{B212}, 451 (1988)}.

\bibitem{Dijkgraaf:1989hb}
R.~Dijkgraaf, C.~Vafa, E.~P. Verlinde \& H.~L. Verlinde,
\textit{``{The Operator Algebra of Orbifold Models}''},
\doiref{10.1007/BF01238812}{Commun.~Math.~Phys. \textbf{123}, 485 (1989)}.

\bibitem{Lunin:2000yv}
O.~Lunin \& S.~D. Mathur,
\textit{``{Correlation functions for M**N / S(N) orbifolds}''},
\doiref{10.1007/s002200100431}{Commun.~Math.~Phys. \textbf{219}, 399 (2001)},
\normalsize{\texttt{\arxivref{hep-th/0006196}{hep-th/0006196}}}.

\bibitem{Belavin:1984vu}
A.~Belavin, A.~M. Polyakov \& A.~Zamolodchikov,
\textit{``{Infinite Conformal Symmetry in Two-Dimensional Quantum Field
  Theory}''},
\doiref{10.1016/0550-3213(84)90052-X}{Nucl.Phys. \textbf{B241}, 333 (1984)}.

\bibitem{Liendo:2012hy}
P.~Liendo, L.~Rastelli \& B.~C. van~Rees,
\textit{``{The Bootstrap Program for Boundary CFT}''},
\normalsize{\texttt{\arxivref{1210.4258}{arXiv:1210.4258}}}.

\bibitem{Gliozzi:2014jsa}
F.~Gliozzi \& A.~Rago,
\textit{``{Critical exponents of the 3d Ising and related models from Conformal
  Bootstrap}''},
\doiref{10.1007/JHEP10(2014)042}{JHEP \textbf{1410}, 042 (2014)},
\normalsize{\texttt{\arxivref{1403.6003}{arXiv:1403.6003}}}.

\bibitem{Gliozzi:2013ysa}
F.~Gliozzi,
\textit{``{More constraining conformal bootstrap}''},
\doiref{10.1103/PhysRevLett.111.161602}{Phys.Rev.Lett. \textbf{111}, 161602
  (2013)},
\normalsize{\texttt{\arxivref{1307.3111}{arXiv:1307.3111}}}.

\bibitem{Teschner1}
J.~Teschner,
\textit{``{A Lecture on the Liouville vertex operators}''},
\doiref{10.1142/S0217751X04020567}{Int.~J.~Mod.~Phys. \textbf{A19S2}, 436
  (2004)},
\normalsize{\texttt{\arxivref{hep-th/0303150}{hep-th/0303150}}},
in \textit{``{Proceedings, 6th International Workshop on Conformal field theory
  and integrable models}''},
p.~436-458.

\bibitem{Teschner2}
B.~Ponsot \& J.~Teschner,
\textit{``{Liouville bootstrap via harmonic analysis on a noncompact quantum
  group}''},
\normalsize{\texttt{\arxivref{hep-th/9911110}{hep-th/9911110}}}.

\bibitem{Teschner3}
J.~Teschner \& G.~Vartanov,
\textit{``{6j symbols for the modular double, quantum hyperbolic geometry, and
  supersymmetric gauge theories}''},
\doiref{10.1007/s11005-014-0684-3}{Lett.~Math.~Phys. \textbf{104}, 527 (2014)},
\normalsize{\texttt{\arxivref{1202.4698}{arXiv:1202.4698}}}.

\bibitem{Dotsenko:1984nm}
V.~S. Dotsenko \& V.~A. Fateev,
\textit{``{Conformal Algebra and Multipoint Correlation Functions in
  Two-Dimensional Statistical Models}''},
\doiref{10.1016/0550-3213(84)90269-4}{Nucl.~Phys. \textbf{B240}, 312 (1984)}.

\bibitem{Dotsenko:1984ad}
V.~S. Dotsenko \& V.~A. Fateev,
\textit{``{Four Point Correlation Functions and the Operator Algebra in the
  Two-Dimensional Conformal Invariant Theories with the Central Charge $c <
  1$}''},
\doiref{10.1016/S0550-3213(85)80004-3}{Nucl.~Phys. \textbf{B251}, 691 (1985)}.

\bibitem{ZamolodchikovRecursion}
A.~Zamolodchikov,
\textit{``{Conformal Symmetry in Two-Dimensions: An Explicit Recurrence Formula
  for the Conformal Partial Wave Amplitude}''},
\doiref{10.1007/BF01214585}{Commun.Math.Phys. \textbf{96}, 419 (1984)}.

\bibitem{Zamolodchikovq}
A.~Zamolodchikov,
\textit{``{Conformal Symmetry in Two-dimensional Space: Recursion
  Representation of the Conformal Block}''},
Teoreticheskaya~i~Matematicheskaya~Fizika \textbf{73}, 103 (1987).

\bibitem{HartmanLargeC}
T.~Hartman,
\textit{``{Entanglement Entropy at Large Central Charge}''},
\normalsize{\texttt{\arxivref{1303.6955}{arXiv:1303.6955}}}.

\bibitem{Gliozzi:2016cmg}
F.~Gliozzi,
\textit{``{Truncatable bootstrap equations in algebraic form and critical
  surface exponents}''},
\normalsize{\texttt{\arxivref{1605.04175}{arXiv:1605.04175}}}.

\bibitem{Gaberdiel:1996kx}
M.~R. Gaberdiel \& H.~G. Kausch,
\textit{``{Indecomposable fusion products}''},
\doiref{10.1016/0550-3213(96)00364-1}{Nucl.~Phys. \textbf{B477}, 293 (1996)},
\normalsize{\texttt{\arxivref{hep-th/9604026}{hep-th/9604026}}}.

\bibitem{Perlmutter:2015iya}
E.~Perlmutter,
\textit{``{Virasoro conformal blocks in closed form}''},
\doiref{10.1007/JHEP08(2015)088}{JHEP \textbf{1508}, 088 (2015)},
\normalsize{\texttt{\arxivref{1502.07742}{arXiv:1502.07742}}}.

\bibitem{Pappadopulo:2012jk}
D.~Pappadopulo, S.~Rychkov, J.~Espin \& R.~Rattazzi,
\textit{``{OPE Convergence in Conformal Field Theory}''},
\normalsize{\texttt{\arxivref{1208.6449}{arXiv:1208.6449}}}.

\bibitem{Mack:1976pa}
G.~Mack,
\textit{``{Convergence of Operator Product Expansions on the Vacuum in
  Conformal Invariant Quantum Field Theory}''},
\doiref{10.1007/BF01609130}{Commun.~Math.~Phys. \textbf{53}, 155 (1977)}.

\bibitem{Rattazzi:2008pe}
R.~Rattazzi, V.~S. Rychkov, E.~Tonni \& A.~Vichi,
\textit{``{Bounding scalar operator dimensions in 4D CFT}''},
\doiref{10.1088/1126-6708/2008/12/031}{JHEP \textbf{0812}, 031 (2008)},
\normalsize{\texttt{\arxivref{0807.0004}{arXiv:0807.0004}}}.

\bibitem{ElShowk:2012hu}
S.~El-Showk \& M.~F. Paulos,
\textit{``{Bootstrapping Conformal Field Theories with the Extremal Functional
  Method}''},
\normalsize{\texttt{\arxivref{1211.2810}{arXiv:1211.2810}}}.

\bibitem{TeschnerLec}
J.~Teschner,
\textit{``{Liouville theory revisited}''},
\doiref{10.1088/0264-9381/18/23/201}{Class.~Quant.~Grav. \textbf{18}, R153
  (2001)},
\normalsize{\texttt{\arxivref{hep-th/0104158}{hep-th/0104158}}}.

\bibitem{DiFrancesco:1997nk}
P.~Di~Francesco, P.~Mathieu \& D.~Senechal,
\textit{``{Conformal Field Theory}''},
Springer-Verlag (1997),
New York.

\bibitem{barnes}
E.~W. {Barnes},
\textit{``{The Theory of the Double Gamma Function}''},
\doiref{10.1098/rsta.1901.0006}{Philosophical~Transactions~of~the~Royal~Society~of~London~Series~A
  \textbf{196}, 265 (1901)}.

\bibitem{Nakayama}
Y.~Nakayama,
\textit{``{Liouville field theory: A Decade after the revolution}''},
\doiref{10.1142/S0217751X04019500}{Int.~J.~Mod.~Phys. \textbf{A19}, 2771
  (2004)},
\normalsize{\texttt{\arxivref{hep-th/0402009}{hep-th/0402009}}}.

\bibitem{spreafico}
M.~Spreafico,
\textit{``On the Barnes double zeta and Gamma functions''},
\doiref{10.1016/j.jnt.2009.03.005}{Journal~of~Number~Theory \textbf{129}, 2035
  (2009)}.

\bibitem{furlan}
P.~Furlan, A.~C. Ganchev \& V.~B. Petkova,
\textit{``{Fusion Matrices and $C < 1$ (Quasi)local Conformal Theories}''},
\doiref{10.1142/S0217751X90001252}{Int.~J.~Mod.~Phys. \textbf{A5}, 2721
  (1990)},
[Erratum: Int. J. Mod. Phys.A5,3641(1990)].

\bibitem{Hou}
B.-Y. Hou, D.-P. Lie \& R.-H. Yue,
\textit{``{Quantum Group Structure in Unitary Minimal Model}''},
\doiref{10.1016/0370-2693(89)90153-6}{Phys.~Lett. \textbf{B229}, 45 (1989)}.

\bibitem{kirillov}
A.~N. {Kirillov} \& N.~Y. {Reshetikhin},
\textit{``{Representations of the algebra $U_q(sl(2))$ , q-orthogonal
  Polynomials and Invariants of Links}''},
\doiref{10.1142/9789812798329_0012}{New~Developments~In~The~Theory~Of~Knots.~Series:~Advanced~Series~in~Mathematical~Physics,~ISBN:~978-981-02-0162-3.~WORLD~SCIENTIFIC,~Edited~by~Toshitake~Kohno,~vol.~11,~pp.~202-256
  \textbf{11}, 202 (1990)}.

\bibitem{paulina-et-al}
M.~Pawelkiewicz, V.~Schomerus \& P.~Suchanek,
\textit{``{The universal Racah-Wigner symbol for U$_q$(osp$(1\vert 2)$)}''},
\doiref{10.1007/JHEP04(2014)079}{JHEP \textbf{1404}, 079 (2014)},
\normalsize{\texttt{\arxivref{1307.6866}{arXiv:1307.6866}}}.

\bibitem{Dotsenko:1985hi}
V.~S. Dotsenko \& V.~A. Fateev,
\textit{``{Operator Algebra of Two-Dimensional Conformal Theories with Central
  Charge $C \leq 1$}''},
\doiref{10.1016/0370-2693(85)90366-1}{Phys.~Lett. \textbf{B154}, 291 (1985)}.

\bibitem{Fuchs}
J.~Fuchs,
\textit{``{Operator Product Coefficients in Nondiagonal Conformal Field
  Theories}''},
\doiref{10.1103/PhysRevLett.62.1705}{Phys.~Rev.~Lett. \textbf{62}, 1705
  (1989)}.

\bibitem{FuchsKlemm}
J.~Fuchs \& A.~Klemm,
\textit{``{The Computation of the Operator Algebra in Nondiagonal Conformal
  Field Theories}''},
\doiref{10.1016/0003-4916(89)90275-3}{Annals~Phys. \textbf{194}, 303 (1989)}.

\end{thebibliography}

\end{document}